\documentclass[openacc]{rsproca_new}%%%%where rsproca is the template name

\usepackage{amssymb,amsmath,placeins,epsfig,array}

\usepackage{physics}
\usepackage{bm}
\usepackage{scalerel}
\usepackage{tikz}
\usepackage{soul}

\usepackage{graphicx}
\graphicspath{ {figs/} }

\usepackage[normalem]{ulem} % This so I can use \sout{}
\newcommand{\markblue}[1]{\textcolor{black}{#1}}

\usepackage{lipsum}

\usepackage{verbatim}

\titlehead{Research}

\begin{document}

%%%% Article title to be placed here
\title{Anomalous interference drives oscillatory dynamics in wave-dressed active particles}

\author{%%%% Author details
Austin M. Blitstein$^{1}$, Rodolfo R. Rosales$^{2}$, Pedro J. S\'{a}enz$^{1}$}

%%%%%%%%% Insert author address here
\address{$^{1}$Department of Mathematics, University of North Carolina, Chapel Hill, North Carolina, 27599, USA\\
$^{2}$Department of Mathematics, Massachusetts Institute of Technology, Cambridge, Massachusetts, 02139, USA}

%%%% Subject entries to be placed here %%%%
\subject{applied mathematics}

%%%% Keyword entries to be placed here %%%%

%%%% Insert corresponding author and its email address}
\corres{Pedro J. S\'{a}enz\\
\email{saenz@unc.edu}}

%%%% Abstract text to be placed here %%%%%%%%%%%%
\begin{abstract}
A recent surge of discoveries has sparked significant interest in active systems where a particle moves autonomously in resonance with its self-generated wave field, leading to notable wave-mediated effects including new propulsion mechanisms, self-sustained speed modulations, and quantum-like phenomena. Drawing from  \markblue{walking droplets,} an archetypical model of wave-dressed active particles, we  \markblue{identify} a wave-mediated non-local force driving their oscillatory dynamics, arising from the particle’s path memory and an unconventional form of wave interference near jerking points, locations where the particle’s velocity changes rapidly. In contrast to the typical case of constructive interference at points of stationary phase, waves excited by the particle near jerking points avoid cancellation through rapid changes in frequency. Through an asymptotic analysis, we derive the wave force from jerking points, revealing it as an \markblue{exponentially small} yet crucial remnant of the particle's past motion. This previously unrecognized force underlies a range of phenomena  \markblue{that have thus far been treated independently}, including in-line speed oscillations, wave-like statistics in potential wells, and non-specular reflections, thereby unifying them within a common framework rooted in generic wave superposition principles.
\end{abstract}
%%%%%%%%%%%%%%%%%%%%%%%%%%%

%\rsbreak

\maketitle

%%%%%%%%%% Insert the texts which can accomdate on firstpage in the tag "fmtext" %%%%%

\section{Introduction}

\begin{figure*}
    \centering
    \includegraphics[width=\linewidth]{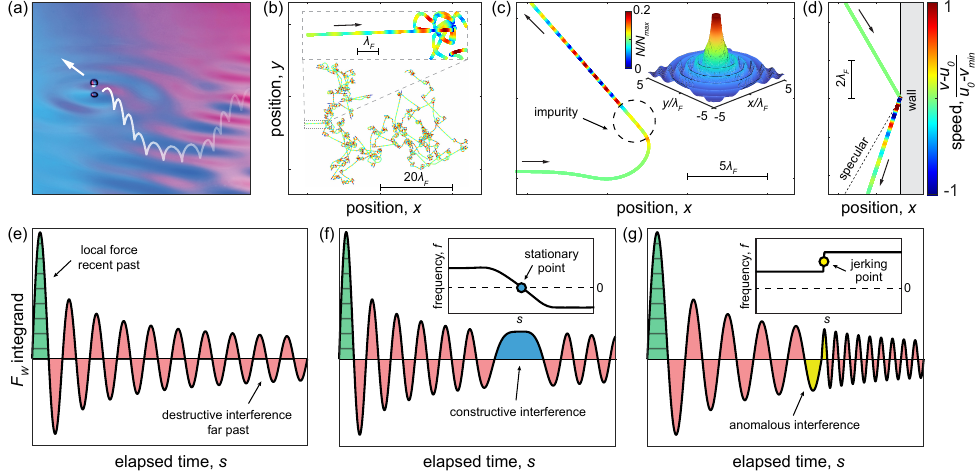}
    \caption{(a) A \markblue{walking droplet} as seen in experiments \cite{couder_2005}. (b-d) By surfing the waves [with wavelength $\lambda_F$] they excite in the past, such \markblue{droplets} exhibit oscillatory dynamics across a range of seemingly disparate phenomena. (b) 
    Spontaneously emerging in-line  speed oscillations \cite{Wind-Willassen2013,bacot2019multistable,durey_2020_speed_osc} [inset] can lead to chaotic two-dimensional motion, exhibiting long-time diffusive behavior \cite{hubert2019tunable,bacot2019multistable,hubert2022overload,durey_2020_hqft,durey_2021,valani_2018,valani_2021,dagan_2020}. 
    (c) Encounters with topographical impurities can likewise trigger these speed oscillations, resulting in a wave-like statistical signature [top right, position histogram of an ensemble of \markblue{droplet} trajectories] reminiscent of the Friedel oscillations exhibited by electrons scattering from impurities \cite{saenz_2020}.         
    (d) Similar oscillatory dynamics may emerge  following the non-specular reflection of a \markblue{droplet} off a wall, even though the initial reflection is  specular [i.e.  the normal component of the velocity is instantaneously reversed at the point of reflection] \cite{nonspecular}. 
    We show that a single underlying mechanism rationalizes all of these distinct phenomena. 
    (e) Schematic of the integrand of the wave force $F_W$ driving a \markblue{droplet} at constant speed along a straight line. The oscillations are the result of the \markblue{droplet} surfing waves it previously emitted, with frequency $f$ representing the rate at which it crosses past wave crests. Waves from the recent past contribute a local force that sets the preferred speed [green \markblue{with horizontal black lines}], while crests and troughs from the distant past cancel due to destructive interference [red].
    (f) Conventionally, this cancellation  may break down at points of stationary phase [blue], where $f = 0$, enabling constructive superposition and resulting in a non-local force \cite{blitstein_2024}. 
    (g) We identify a qualitatively different disruption of wave cancellation due to anomalous interference at jerking points [yellow], where the wave surfing frequency $f$ changes rapidly. This sudden shift induces a mismatch in size between consecutive peaks and troughs in the integrand near jerking points, in contrast with the broadening of oscillations seen near stationary points. Jerking points naturally generate cascades of non-local oscillatory forces that unify the diverse oscillatory behaviors described in (b-d).}
    \label{fig:1}
\end{figure*}

Wave-dressed active particles are gaining attention at an increasing rate, initially driven by the discovery of self-propelled walking droplets that mimic quantum phenomena [Fig.\,\ref{fig:1}(a)] \cite{couder_2005,bush_2015,bush_2020,bush2024perspectives}, and more recently accelerated by the realization of various analogous active systems exhibiting dual wave-particle features \cite{honeybees,gunwale,benham2024wave,longuet1977mean,surfer,tarr_2024,ludion,acoustic,acoustic2,surferbot,dagan_2020,durey_2020_hqft}. The particle undergoes intrinsic oscillations that excite waves in the surrounding field, which in turn feed back onto the particle and influence its motion. This class of wave-mediated dynamics has now been observed across a wide range of fields and scales, including insects using capillary waves for survival \cite{honeybees}, people or robots surfing self-generated surface waves \cite{gunwale,benham2024wave,longuet1977mean,surfer,tarr_2024,surferbot}, submersibles navigating via internal waves \cite{ludion}, and bubbles propelled by their own acoustic fields \cite{acoustic,acoustic2}. Despite its ubiquity, developing a general mechanistic framework remains challenging. These systems operate far from equilibrium, where familiar conservation laws do not apply directly \cite{ramaswamy2010mechanics,marchetti2013hydrodynamics,julicher2018hydrodynamic}; their dynamics span disparate timescales, with internal oscillations typically much faster than the particle's long-term translation \cite{harris_2013_b,Saenz2018,cristea_2018,durey_2020_corral,abraham2024classical}; and they are further enriched by the effects of path memory \cite{fort_2010,eddi_2011,molavcek_2013,oza_2013,tadrist2018faraday}. 
Unlike other active particles with memory effects that may be either `self-seeking', such as chemotactic bacteria \cite{budrene_1995,mittal_2003}, or `self-avoiding', such as swimming oil droplets \cite{maass2016swimming,jin2017chemotaxis,billiards}, wave-mediated forces oscillate in space -- switching between attraction and repulsion -- and can thus intermittently cancel along significant portions of the trajectory. 
These complexities make the search for unifying principles both difficult and rewarding, whether the goal is to understand evolutionary survival strategies, inform robotic design, develop macroscale quantum analogs \cite{bush_2015,bush_2020,bush2024perspectives}, or uncover mechanisms for wave-mediated collective order \cite{Saenz2021,surfer}.

Here, we show that a range of distinctive behaviors exhibited by walking droplets, a canonical example of wave-dressed active particles, can all be traced to a single form of wave interference that has gone unrecognized until now. These include sustained speed oscillations~\cite{Wind-Willassen2013,bacot2019multistable,durey_2020_speed_osc}, stochastic transitions at high path memory~\cite{hubert2019tunable,bacot2019multistable,hubert2022overload,durey_2020_hqft,durey_2021,valani_2018,valani_2021,dagan_2020}, quantum-like statistics around impurities~\cite{saenz_2020} and within potential wells~\cite{durey2018dynamics,montes_2019,montes_2021}, and non-specular boundary reflections~\cite{nonspecular} [Fig.~\ref{fig:1}(b–d)].
In contrast to the traditional case of constructive interference, where waves add coherently in phase, we \markblue{show} that these phenomena arise from an \textit{anomalous} type of interference in which waves along the path avoid cancellation due to an abrupt change in the frequency at which the \markblue{droplet} surfs previously emitted waves. Through asymptotic analysis, we demonstrate that this anomalous interference generates spatially oscillatory, non-local forces \markblue{[i.e.~forces that depend, via persistent waves, on variables at past locations and times]} with the same wavelength as the underlying wave field, localized at points along the trajectory where the \markblue{droplet's} velocity changes rapidly. Though subtle, this wave-mediated force plays a key role in driving and statistically shaping the oscillatory dynamics that set \markblue{walking droplets} apart.

At the core of the anomalous interference mechanism we identify is the concept of \textit{jerking points}, discrete locations along the \markblue{droplet}'s past trajectory from which we show a wave-mediated non-local force emanates. When the \markblue{droplet} moves at constant velocity, the waves it excited along its past trajectory destructively interfere at its current location, except for the most recent waves, which have not yet canceled [Fig.\,\ref{fig:1}(e)]. As a result, there is no force originating from the distant past, while the recent waves generate a local force that drives an overdamped relaxation toward a preferred speed \cite{bush_2014,labousse2014non}, which is known to be insufficient to produce wave-like particle dynamics \markblue{[i.e. particle motion reminiscent of waves, such as oscillatory spatial histograms and speed oscillations with a well-defined spatial wavelength]}. The standard mechanism by which waves emitted in the \markblue{distant} past can avoid cancellation is constructive interference at stationary points \cite{fort_2010,perrard2014self,perrard2014chaos,labousse2014build,oza_2014_a,oza_2014_b,Labousse_2016,blitstein_2024}, where the \markblue{droplet} briefly stops or maintains a fixed distance from earlier positions, corresponding to points of stationary phase in the wave force integrand [Fig.\,\ref{fig:1}(f)]. Owing to the associated widening of oscillations in the integrand, stationary points produce a spatio-temporally non-local force \cite{blitstein_2024} that has been shown to underlie quantized orbits and preferred path curvatures \cite{fort_2010,perrard2014self,perrard2014chaos,labousse2014build}, but it too cannot account for the \markblue{droplet's} self-sustained oscillatory dynamics and quantum-like statistics \cite{Wind-Willassen2013,bacot2019multistable,durey_2020_speed_osc,saenz_2020,durey2018dynamics,montes_2019,montes_2021,nonspecular}.

The anomalous interference at jerking points revealed here circumvents the cancellation of waves emitted in the \markblue{distant} past in a fundamentally different manner: it arises from rapid changes in the \markblue{droplet's} velocity, which shift the frequency of wave surfing, thus leading to a size mismatch between consecutive peaks and troughs in the wave-force kernel [Fig.\,\ref{fig:1}(g)]. We derive an asymptotic approximation for the non-local force resulting from this frequency mismatch and show that the associated anomalous interference may propagate through a cascade of self-excited jerking points, inducing in-line speed oscillations \cite{Wind-Willassen2013,bacot2019multistable,durey_2020_speed_osc} that underlie key hydrodynamic quantum analogs \cite{saenz_2020,durey2018dynamics,montes_2019,montes_2021} and stochastic dynamics \cite{hubert2019tunable,hubert2022overload,durey_2020_hqft,durey_2021,valani_2018,valani_2021,dagan_2020}, and in two dimensions non-specular wall reflections \cite{nonspecular}. Our analysis shows that the jerking-point force is exponentially weaker than the local one, highlighting its elusiveness, yet it is the essential ingredient that unifies these \markblue{behaviors that have thus far been treated independently} under a common mechanistic framework.

\section{Model Framework}

\markblue{Consider a walking droplet on the surface of a vertically vibrated bath of the same fluid, of constant depth and effectively infinite horizontal extent \cite{couder_2005,bush_2015,bush_2020,bush2024perspectives,eddi_2011,molavcek_2013,oza_2013,tadrist2018faraday}. The bath oscillates with acceleration $\gamma\cos(2\pi f_B t)$, where $\gamma$ is the amplitude of the acceleration and $f_B$ the forcing frequency. The dynamics of interest arise when $\gamma$ is kept below, but close, to the Faraday threshold $\gamma_F$, above which subharmonic, monochromatic standing waves with wavenumber $k_F = 2\pi/\lambda_F$ and period $T_F = 2/f_B$ are parametrically excited \cite{faraday_1831,kumar_tuckerman_1994}. In this regime, a droplet placed on the free surface may bounce indefinitely, and each bounce produces a localized Faraday wave that decays exponentially with a characteristic decay, or `memory', time $T_M = T_d/(1-\gamma/\gamma_F)$, where $T_d$ is the characteristic decay time of the waves in the absence of vibration \cite{molavcek_2013}. As $\gamma\rightarrow\gamma_F$, the system `path memory' increases \cite{fort_2010}: as the waves persist longer in the bath, the droplet dynamics are influenced by an increasingly long past trajectory \cite{eddi_2011}. 
Following \cite{oza_2013}, we neglect the far-field spatial decay, which  provides only small corrections to the dynamics considered here, so that the wave field produced at each bounce is proportional to a zeroth-order Bessel function centered at the droplet. The effects we investigate arise when the droplet is a resonant walker, meaning that its bouncing period is twice that of the bath, $1/f_B$, and thus matches the Faraday period $T_F$, while the impact phase relative to the wave field  remains nearly constant in time \cite{oza_2013}. At each bounce, the bath exerts a horizontal force on the droplet proportional to the negative gradient of the wave field at the impact location \cite{molavcek_2013}. Assuming the droplet translates sufficiently slowly in the horizontal direction relative to the bouncing frequency, the forces acting on the droplet may be averaged over the bouncing period, yielding an integro-differential equation of motion \cite{oza_2013} for the droplet position $\bm{x}_p(t)$ in the plane of the bath:}
\begin{equation}\label{eq:eom1}
    m\ddot{\bm{x}}_p + D\dot{\bm{x}}_p = \bm{F}_W + \bm{F}.
\end{equation}
\markblue{Here, dots denote differentiation with respect to time, $m$ is the droplet mass, $D$ is the linear drag coefficient capturing dissipation during flight and impact with the bath, $\bm{F}$ is some yet to be specified external force acting on the droplet without directly influencing the underlying wave field $h(t,\bm{x})$, and $\bm{F}_W$ is the wave force}
\begin{equation}\label{eq:eom2}
    \bm{F}_W=-mg\mathcal{S}\nabla h(t,\bm{x}_p(t))=-\frac{mg\mathcal{S}A}{T_F}\nabla\int_{-\infty}^{t}J_0\left(k_F|\bm{x}_p(t)-\bm{x}_p(t')|\right)e^{-(t-t')/T_M}dt',
\end{equation}
\markblue{where $g$ is the gravitational acceleration, $\mathcal{S}$ is the sine of the impact phase, and $A$ is the wave amplitude [shown in \cite{oza_2013} to depend on the droplet radius, gravity, the Faraday wavelength and period, and the fluid surface tension, viscosity, and density].} 
%We discuss later how the theoretical analysis that follows may be modified when the assumptions underlying \eqref{eq:eom1}-\eqref{eq:eom2} are relaxed, both in the context of walking droplets\cite{galeano2018ratcheting,couchman2019bouncing,couchman2020free,primkulov2025diffraction,primkulov2025nonresonant,evans2026phase} and in extending to other wave-dressed active particle systems \cite{honeybees,gunwale,benham2024wave,longuet1977mean,surfer,tarr_2024,ludion,acoustic,acoustic2,surferbot,dagan_2020,durey_2020_hqft}.} 
%and note for now that the explicit formulas presented here are only appropriate in so far as the equations of motion hold true.}

Following \cite{Oza_2018,durey_2020_speed_osc}, we rewrite \eqref{eq:eom1}-\eqref{eq:eom2} in dimensionless form as
\begin{align} \kappa_0\ddot{\bm{x}}_p+\dot{\bm{x}}_p=\bm{F}_W+\bm{F}, \label{eq:eom3} \\
\bm{F}_W = \int_0^\infty 2J_1(d(s))\hat{\bm{d}}(s)e^{-(1-\Gamma)s}ds. \label{eq:eom4}
\end{align}
\markblue{Here, position is nondimensionalized by the inverse wavenumber $k_F^{-1}$ and time by $T_M(1-\Gamma)$, where $\Gamma=0$ corresponds to the walking threshold $\gamma=\gamma_W$ and $\Gamma=1$ to the Faraday threshold $\gamma=\gamma_F$ [$\Gamma = (\gamma-\gamma_W)/(\gamma_F-\gamma_W)$ with $\gamma_W/\gamma_F = 1 - \sqrt{(mgA\mathcal{S}k_F^2T_d^2)/(2DT_F)}$], and $\kappa_0 = m/(DT_M(1-\Gamma))$ is the dimensionless mass. For convenience, we have rewritten the integral of the droplet's past history in terms of the elapsed time $s$, and introduced the notation $\bm{d}(s)=\bm{x}_p(t)-\bm{x}_p(t-s)$ for the displacement vector connecting a past droplet location $\bm{x}_p(t-s)$ to the current position $\bm{x}_p(t)$, with magnitude $d(s)=|\bm{d}(s)|$ and direction $\hat{\bm{d}}(s)=\bm{d}(s)/d(s)$. While we  suppress the explicit $t$-dependence in $\bm{d}(s)$ to avoid notational clutter, we emphasize that any function of $\bm{d}(s)$ evolves in time $t$, including the wave-surfing frequency and the location of the jerking points introduced below. In what follows, all variables are expressed in these dimensionless units [note that we reuse some symbols from the dimensional equations], and dots and primes denote [partial] differentiation with respect to  dimensionless time $t$ and elapsed time $s$, respectively.}

\begin{figure}
    \centering
    \includegraphics[width=\linewidth]{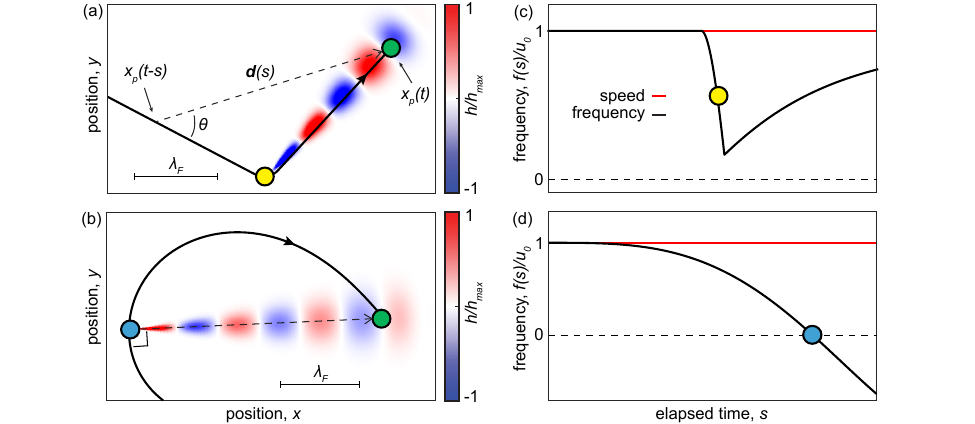}
    \caption{(a) Jerking points [yellow] may arise geometrically [without a change in speed] when the angle $\theta$ between the displacement vector and trajectory changes rapidly, to be compared with (b) stationary points [blue] that arise geometrically when $\theta = 90^{\circ}$ \cite{blitstein_2024}. In both cases, the only waves $h$ in the distant past that avoid cancellation at the \markblue{droplet's} current position [green] are from the jerking and stationary points. \markblue{(c-d) Corresponding wave-surfing frequency $f$ for (a) and (b), respectively. For sake of illustration, the droplet speed here is held constant, so that variations in $f$ arise solely from the $\cos\theta(s)$ factor in \eqref{eq:frequency}. In the remainder of the study, the speed evolves dynamically according to \eqref{eq:eom3}-\eqref{eq:eom4}.}}
    \label{fig:2}
\end{figure}

Since the spatial profile of the wave-force kernel, \markblue{$J_1(d)$}, asymptotes to a sinusoid with slowly changing amplitude for large arguments \cite{bender_orszag_1978}, as $d\rightarrow \infty$ [in practice $d\gtrsim O(1)$], there is an associated phase [given by the argument of the sinusoid up to an additive constant] equal to $\varphi(s)=d(s)$. Central to our investigation is the associated frequency $f(s)=\varphi'(s)=d'(s)$, which may be expressed as
\begin{equation}\label{eq:frequency}
f(s) = \bm{\dot{x}}_p(t - s)\cdot \bm{\hat{d}}(s) =|\bm{\dot{x}}_p(t-s)|\cos\theta(s),
\end{equation}
where $\theta(s)$ is the angle between the velocity at a past point, $\bm{\dot{x}}_p(t-s)$, and the displacement vector, $\bm{d}(s)$  [Fig.\,\ref{fig:2}(a)]. \markblue{We refer to $f(s)$ as the wave-surfing frequency, since \eqref{eq:frequency} represents the instantaneous rate at which the droplet traverses} the waves it previously excited along its path. \markblue{For a droplet moving at constant speed $u_0$ along a straight path, the surfing frequency reduces to $f = u_0$ [i.e.~speed times wavenumber in dimensional units]. In general, however, for curved motion or variable velocity, the surfing frequency depends on both the droplet’s velocity and its past trajectory, as reflected in the projection in \eqref{eq:frequency}.
}
%\markblue{and should not be confused with the frequency of bath vibration $f_B$ nor the frequency of the waves the droplet excites $1/T_F=f_B/2$.}

\section{Anomalous Interference}

To illustrate the physical and mathematical structure underlying anomalous interference, we begin by analyzing the simpler case of \markblue{rectilinear motion before generalizing the results to curved trajectories. We first present the key ideas at a conceptual level and defer the full mathematical formulation to Section \ref{sec:asymptotic_approx}.}.

\subsection{One-Dimensional Motion}

While the \markblue{droplet} moves at constant speed along a straight path, it generates \markblue{radially symmetric} standing waves uniformly along its trajectory -- waves that it must later traverse. Because the source is moving, these standing waves eventually become out of phase in space, resulting in destructive interference at the \markblue{droplet's} current location \cite{eddi_2011,molavcek_2013,oza_2013}. Mathematically, this physical picture translates to an integral kernel for the wave force, \eqref{eq:eom4}, that oscillates with the distance [or equivalently, the time] to past locations where the waves were emitted [Fig.\,\ref{fig:1}(e)]. In \markblue{rectilinear motion}, the oscillation frequency \eqref{eq:frequency} reduces to $f(s) =\pm|\bm{\dot{x}}_p(t-s)|$, with the sign distinguishing motion towards or away from past wave emission points [$\cos\theta(s)=\pm1$].
As the \markblue{droplet} surfs over past waves at this rate, consecutive peaks and troughs cancel one another, yielding a negligible force from the distant past [Fig.\,\ref{fig:1}(e), red]. Typically, the only exception to this destructive interference occurs near the \markblue{droplet's} current location [Fig.\,\ref{fig:1}(e), green], where recently generated waves have not yet canceled [since waves have been produced only on one side of the \markblue{droplet}, along its recent trajectory], resulting in a local force [i.e. dependent solely on variables at the current location and time] that serves to set the \markblue{droplet's} preferred speed. This interference mechanism is well documented, and it has been established that the resulting local force \cite{bush_2014} cannot account for the speed oscillations \cite{Wind-Willassen2013,bacot2019multistable,saenz_2020,durey_2020_speed_osc} underpinning a range of phenomena in  walking droplets \cite{bush_2015,bush_2020,bush2024perspectives}.

For the distinctive dynamics of \markblue{walking droplets captured by the integro-differential model \eqref{eq:eom3}-\eqref{eq:eom4}} to emerge, the previous destructive interference must be disrupted so that waves emitted in the distant past can generate spatio-temporally non-local forces at the \markblue{droplet's} position [i.e.~forces that depend, via persistent waves, on variables at past locations and times] \cite{blitstein_2024}. \markblue{This apparent nonlocality arises from eliminating the underlying wave field in the equations of motion. At the fundamental level, the \markblue{droplet} interacts locally with the wave it encounters, while the wave itself carries information forward in space and time. In this sense, the nonlocality thus refers to the effective, history-dependent structure of the reduced equations of motion.}

The typical mechanism by which such a disturbance of destructive interference can occur is through constructive interference at points of stationary phase. Physically, in one-dimensional motion, this occurs when the \markblue{droplet} momentarily stops, thereby exciting more waves at a particular location and creating a local buildup of wave amplitude. Mathematically, the stopping location corresponds to a point of stationary phase \cite{blitstein_2024}, where the frequency of the wave kernel \eqref{eq:frequency} vanishes since $|\bm{\dot{x}}_p(t-s)|=0$, and thus a particular peak or trough broadens [Fig.\,\ref{fig:1}(f), blue]. This mechanism dates back to Lord Rayleigh's work on wave propagation and asymptotics  \cite{Rayleigh}, and classical techniques such as the method of stationary phase and the method of steepest descent \cite{copson_1965} are routinely used to extract contributions from such points. In the context of walking droplets, this mechanism is well documented, and the non-local force emanating from stationary points \cite{blitstein_2024} has been shown to underlie orbital quantization in the presence of Coriolis forcing \cite{fort_2010,oza_2014_a,harris2014droplets}. However, the stationary-point force does not account for the oscillatory dynamics observed in walking droplets \cite{Wind-Willassen2013,bacot2019multistable,durey_2020_speed_osc,saenz_2020,durey2018dynamics,montes_2019,montes_2021,hubert2019tunable,hubert2022overload,durey_2020_hqft,durey_2021,valani_2018,valani_2021,dagan_2020}, pointing to the existence of additional non-local forces \markblue{whose origin and explicit form have not been identified}.

We shall demonstrate mathematically that the \markblue{missing} force arises from a different and more unconventional mechanism for disrupting the destructive interference of distant waves, yielding a qualitatively distinct type of non-local force. Its physical origin lies at points where the \markblue{droplet} changes velocity rapidly [which we refer to as jerking points] due to an interference mismatch arising from the abrupt change in the rate at which the \markblue{droplet} surfs the waves it laid down along its path. Mathematically, such rapid variation in \markblue{droplet} velocity leads to a sharp shift in the frequency of the wave kernel \eqref{eq:frequency} at jerking points, in contrast to the vanishing frequency that characterizes stationary points. Rather than broadening oscillations, as in the stationary-phase scenario, this frequency shift induces a local size mismatch between consecutive peaks and troughs in the oscillatory kernel [Fig.\,\ref{fig:1}(g), yellow], leading to a fundamentally different type of non-local force. Since this breakdown of destructive interference differs from the classical stationary-phase mechanism and, to our knowledge, lacks established asymptotic techniques to systematically quantify it, we refer to it as anomalous interference. 

\subsection{Two-Dimensional Motion}

\markblue{When the droplet trajectory departs from rectilinear motion}, both jerking points [associated with anomalous interference] and stationary points [associated with standard constructive interference] may arise not only from changes in speed but also from geometric features of the trajectory, even when the \markblue{droplet} moves at constant speed [Fig.\,\ref{fig:2}]. While the mathematical definitions of stationary points [$f(s)$ vanishes] and jerking points [$f(s)$ varies rapidly] remain formally unchanged, the generalized definition of the wave-surfing frequency \eqref{eq:frequency} shows that the rate at which the \markblue{droplet} traverses previously emitted waves can either vanish or vary sharply due solely to geometric changes in the trajectory through $\theta(s)$.
This geometric dependence leads to richer physical implications in two-dimensional motion.

%Stationary points have been known to arise even if the droplet does not come to rest [Fig.\,\ref{fig:2}(b)], with those appearing in constant-speed circular orbits providing the most prominent example \cite{blitstein_2024}. Similarly, 
%Specifically, we demonstrate that jerking points may arise without a rapid change in speed [Fig.\,\ref{fig:2}(a)], a case illustrated by the reflection of a droplet from a potential wall [Fig.\,\ref{fig:5}]. Moreover, while they primarily act along the tangential direction, thereby triggering in-line speed oscillations, geometrically induced jerking points can also exert significant lateral forces, giving rise to phenomena that may have not initially appeared related. Our wall-reflection investigation, for instance, shows that the jerking-point force renders the reflection non-specular [Fig.\,\ref{fig:5}]. 
%For completeness, we note that stationary points have also been shown to arise from the geometry of the trajectory, even when the droplet does not come to rest [Fig.\,\ref{fig:2}(b)], with constant-speed circular orbits providing the most prominent example \cite{blitstein_2024}. 
%For general curved trajectories, it also becomes apparent that the force associated with jerking points acts predominantly along the tangential direction, while the stationary-point force acts primarily in the normal direction, emphasizing their complementary roles [Fig.\,\ref{fig:2}].

\markblue{Specifically, we consider the reflection of a droplet from a potential wall [Fig.\,\ref{fig:5}] to illustrate the geometric origin of jerking points and their subtle dynamical influence. In this case, jerking points arise without a change in speed, generating a wave force that primarily acts along the tangential direction [Fig.\,\ref{fig:2}(a)], which drives the in-line speed oscillations observed in the outgoing trajectory. At the same time, the fact that the reflection angle exceeds the incident angle demonstrates that jerking points can also generate significant lateral deflections, giving rise to behaviors that may not at first appear to share a common origin with speed oscillations, such as non-specular reflection.
For completeness, we note that stationary points have also been shown to arise from the geometry of the trajectory, even when the droplet does not come to rest [Fig.\,\ref{fig:2}(b)], with constant-speed circular orbits providing the most prominent example \cite{blitstein_2024}.
Taken together, these observations reveal a clear geometric decomposition of the wave-induced force: jerking-point contributions act predominantly tangentially, while stationary-point contributions act primarily in the normal direction, emphasizing their complementary roles [Fig.\,\ref{fig:2}].}

\section{Asymptotic Approximations\label{sec:asymptotic_approx}}

To demonstrate mathematically that anomalous interference at jerking points drives oscillatory dynamics in \markblue{walking droplets}, we seek to asymptotically reduce the wave force $\bm{F}_W$ \eqref{eq:eom4} into a non-local contribution from jerking points, $\bm{F}_N$, which embeds the system's preferred length scale $\lambda_F$ in the \markblue{droplet's} dynamics, and a local contribution from the \markblue{droplet's} recent history, $\bm{F}_L$, which drives the \markblue{droplet} to a preferred speed $u_0$. 

\subsection{Jerking Points}

The location $\bm{x}_p(t-s_j)$ is identified as a jerking point if the surfing frequency with which the \markblue{droplet} encounters waves from its past \eqref{eq:frequency} undergoes a rapid change [relative to the local oscillation period $2\pi/f$] at $s_j$, which we asymptotically approximate as an instantaneous transition between two constant values [Fig.\,\ref{fig:1}(g)]. More precisely, if $\Delta t_j$ denotes the duration of this transition, $\bm{x}_p(t-s_j)$ is a jerking point if $f\Delta t_j \lesssim O(1)$, though we formally invoke the limit $f\Delta t_j \to 0$. In this limit, the precise location of the jerking point within the interval $\Delta t_j$ does not matter, though we later specify a particular choice that exhibits the best numerical agreement when $f\Delta t_j\lesssim O(1)$. Here, the subscript $j$ denotes the $j$-th most recent jerking point. \markblue{We note that the term ``jerking point'' reflects the large jerks [i.e. changes in acceleration] as the droplet enters and exits the rapid change in velocity.}

Due to the exponential decay in the wave force integral \eqref{eq:eom4}, `standard' asymptotic approximations \cite{copson_1965,bender_orszag_1978} would neglect the influence of jerking points, only capturing the contributions from the \markblue{droplet's} recent past [where $s$ is small]. However, our study demonstrates that, despite not being of leading order, the contributions from jerking points play a crucial role in the dynamics of \markblue{walking droplets}.
%, giving rise to some of the unique dynamics of wave-dressed active droplets. 
We thus apply `physically informed' asymptotics, a strategy that is typical of situations where contributions beyond all orders matter \cite{beyond_all_orders_2012}, and which was successfully used to derive the force from stationary points \cite{blitstein_2024}.

To begin, consider an idealized trajectory where the change in frequency $f(s)=d'(s)$ \eqref{eq:frequency} at each jerking point $s_j$ is both instantaneous and between two constant values, the most recent, $d'_{j,a}$, and the following, $d'_{j,b}$, as in Fig.\,\ref{fig:1}(g). We assume each jerking point is sufficiently far from the \markblue{droplet's} current location [$d(s_j) \gg 1$] so that we may use the large argument asymptotic expansion $2J_1(d)\sim b(d)\cos(d+\phi)$, where $b(d)=2\sqrt{2/\pi d}$ is slowly varying [$\frac{1}{b}\dv{b}{d} \ll 1$] and $\phi=-3\pi/4$ constant \cite{bender_orszag_1978}. \markblue{We note that the asymptotic form of the wave kernel as a sinusoid with slowly varying amplitude holds in any number of spatial dimensions under the assumptions underlying \eqref{eq:eom1}-\eqref{eq:eom4}. Accordingly, our results are readily generalizable beyond walking droplets on a two-dimensional fluid bath [see \cite{SM}].}
%, and treat $\bm{\hat{d}}(s)$ as slowly varying [since $\bm{\hat{d}}'(s)=O\left(1/d(s)\right)$ \cite{SM}].

The anomalous lack of wave cancellation at the \markblue{droplet's} current location arises from the frequency mismatch at each jerking point $s_j$. We thus split the wave force integral \eqref{eq:eom4} into sub-intervals $[s_j,s_{j+1}]$ [the first being $[0,s_1]$] to accommodate the piecewise discontinuities in $d'(s)$ near each jerking point,
\begin{equation}\label{eq:dp_ideal}
    d'(s)=
    \begin{cases}
        d'_{j,a} & s \leq s_j, \\
        d'_{j,b} & s > s_j.
    \end{cases}
\end{equation}
Upon substituting the asymptotic expansion for $2J_1(d)$, the wave force integral \eqref{eq:eom4} may thus be written as
\begin{equation}\label{eq:FN}
    \bm{F}_W \sim \real\left[e^{i\phi}\sum_j\int_{s_j}^{s_{j+1}}b(d(s))\hat{\bm{d}}(s)e^{id(s)-(1-\Gamma)s}\dd s\right].
\end{equation}
Within each sub-interval, the frequency $d'(s)$, local magnitude $b(d(s))$, and direction $\hat{\bm{d}}(s)$ are all slowly varying in the limit that $d(s)$ is large and the \markblue{droplet} speed varies slowly around the preferred speed $u_0$ away from jerking points [see Appendix \ref{sec:AppendixA} for more details].
Under these conditions, and assuming no stationary points [treated in \cite{blitstein_2024}], integration by parts [integrating the exponential \cite{copson_1965}] shows that the leading-order contribution in the large-distance asymptotic limit arises from the boundary terms at the endpoints of the sub-intervals [see Appendix \ref{sec:AppendixA}].
The force associated with each jerking point $s_j$ can thus be understood as the sum of the right-endpoint contribution from the sub-interval $[s_{j-1},s_j]$ and the left-endpoint contribution from the sub-interval $[s_j,s_{j+1}]$. Grouping the endpoint contributions for each $s_j$ in this manner yields the wave force from jerking points 
\begin{equation}\label{eq:jp1}
    \bm{F}_N \sim \sum_j 2D_jJ_1(d(s_j)+\theta_j)e^{-(1-\Gamma)s_j}\bm{\hat{d}}(s_j),
\end{equation}
where $D_j$ and $\theta_j$ are the magnitude and argument of
\begin{equation}\label{eq:jp2}
    \frac{1}{id'_{j,a}-(1-\Gamma)} - \frac{1}{id'_{j,b}-(1-\Gamma)}=D_je^{i\theta_j}.
\end{equation}
While our calculation formally yields a force factor $b(d(s_j))\cos(d(s_j)+\theta_j+\phi)$ [valid for $d(s_j)\gg1$], we have replaced it in \eqref{eq:jp1} with $2J_1(d(s_j)+\theta_j)$ for better % asymptotic
agreement at small $d(s_j)$. 
%\markblue{This enables us to use \eqref{eq:jp1}-\eqref{eq:jp2} even for the most recent jerking point, where $s_j$ and $d(s_j)$ are small [albeit occasionally at the expense of a slight phase shift in the speed oscillations as in Fig.~\ref{fig:5}], despite this regime formally lying outside the asymptotic limits used to derive the jerking-point force.}
\markblue{This replacement allows \eqref{eq:jp1}–\eqref{eq:jp2} to be applied even to the most recent jerking point, where $s_j$ and $d(s_j)$ are small, although this regime formally lies outside the asymptotic limits used in the derivation. In practice, this introduces only a slight phase shift in the resulting speed oscillations [see Fig.~\ref{fig:5}].}
Additional limitations and modifications of our asymptotic derivation, which breaks down for slowly changing frequencies and when both stationary and jerking points coincide, are discussed in \cite{SM}.
We also note that the endpoint contribution near $s=0$ [first sub-interval] must be treated separately, since 
$d(s)$ becomes infinitesimally small there and the large-argument asymptotic expansion therefore breaks down beyond the reach of the above procedure [this special case is addressed in section \ref{sec:local_force}]. By contrast, the endpoint contribution as $s \to \infty$ is exponentially suppressed.

\subsection{Non-Idealized Trajectory}

For trajectories where the kernel frequency \eqref{eq:frequency} changes continuously rather than instantaneously, we may still use the results from the preceding section provided the change in frequency is sufficiently rapid. To do so, we approximate each change in frequency to leading order as being piecewise discontinuous, choosing two constant frequency values and identifying the point within the continuous change at which the approximate discontinuity should be placed. Note that the location of the instantaneous change in frequency for this approximation is what we refer to as a jerking point for a non-idealized trajectory.

We rely on geometrical reasoning to deduce the best approach to approximate each change in frequency in the form of \eqref{eq:dp_ideal}. First, the only places where $d'(s)$ is approximately constant are near its extrema where $d''(s)=0$. Thus, for each subsequent pair of extrema, the most recent at $s_{j,a}$ and the following at $s_{j,b}$ with $s_{j,a}\leq s_{j,b}$, we assign $d'_{j,a}=d'(s_{j,a})$ and $d'_{j,b}=d'(s_{j,b})$ so that \eqref{eq:dp_ideal} agrees with first order Taylor approximations at both extrema. We then define the jerking point $s_j$ such that there is equal area between the actual frequency $d'(s)$ and our piecewise approximation \eqref{eq:dp_ideal} on either side of $s_j$ [i.e. $\int_{s_{j,a}}^{s_j}|d'(s)-d'_{j,a}| = \int_{s_j}^{s_{j,b}}|d'(s)-d'_{j,b}|$], since this balances the error of the approximation on both sides of the jerking point.
%inflection point of the change between both extrema, where $d'''(s_j)=0$, since this places the instantaneous change where $d'(s)$ is changing most rapidly. 
We may then use \eqref{eq:jp1} to calculate the jerking point force for a non-idealized trajectory [to leading order] in the limit that the change in frequency is instantaneous and between two constant values. Alternative definitions of the jerking point, such as the inflection point [$d'''(s_j)=0$] or the midpoint [$s_j = (s_{j,a}+s_{j,b})/2$] of the change, are also possible. However, we find that choosing $s_j$ to balance the areas on both sides of $s_j$ yields better agreement in the simulations presented below.

%Though we predominantly use the balance of areas to find the location of the jerking point, $s_j$, we occasionally find better agreement with \eqref{eq:eom4} when we take $s_j$ to be the midpoint of the extrema, $(s_{j,a}+s_{j,b})/2$. This is typically the case only during the creation of a new jerking points [i.e.~for the most recent jerking point], as it may happen that the true inflection point has yet to be created, thus incorrectly placing the jerking point at the particle's current location where $|d''(s)|$ is temporarily largest. For the simulations to follow, we thus use the midpoint for the most recent jerking point and inflection point for all other jerking points.

\subsection{Local Force}\label{sec:local_force}

The endpoint contribution near $s = 0$ [first sub-interval] corresponds to the local force arising from the recent history, which sets the \markblue{droplet's} preferred speed. This force may be calculated by expanding the magnitude $d(s)\sim |\dot{\bm{x}}_p|s$ and direction $\hat{\bm{d}}(s)\sim\dot{\bm{x}}_p/|\dot{\bm{x}}_p|$ of the displacement in the recent past, near $s=0$, yielding
\begin{equation}\label{eq:local}
    \bm{F}_L \sim \frac{\dot{\bm{x}}_p}{|\dot{\bm{x}}_p|}\int_0^\infty 2J_1(|\dot{\bm{x}}_p|s)e^{-\alpha s}\dd s,
\end{equation}
where $\dot{\bm{x}}_p$ is evaluated at the current time $t$. The balance between $\bm{F}_L$ and $\bm{\dot{x}}_p$ in \eqref{eq:eom3} drives the walking \markblue{droplet} to a preferred walking speed $u_0=\sqrt{4-(1-\Gamma)^2-(1-\Gamma)\sqrt{(1-\Gamma)^2+8}}/\sqrt{2}$, as shown in \cite{oza_2013}.

Note that the integral in \eqref{eq:local} may be computed in closed form, and is equivalent to the leading-order force on the \markblue{droplet} in the weak-acceleration limit, $\dot{\bm{x}}_p(t) = \bm{v}(\epsilon t)$, derived by Bush \textit{et~al.}~\cite{bush_2014}. We neglect here the next-to-leading-order \markblue{$O(\epsilon)$ correction, 
also derived in \cite{bush_2014} and capturing the wave-induced added mass of the droplet}, since we find that the force from the most recent jerking point \markblue{already accounts for this effect}. This overlap arises because the most recent jerking point depends on the change in velocity between the location of the most recent extrema of $d'(s)$ and its current position, which provides a measure of the \markblue{droplet's} current acceleration $\ddot{x}_p(t) = \epsilon\dot{\bm{v}}(\epsilon t)$ appearing at $O(\epsilon)$.

\subsection{Minimal Model}

Combining the contributions from jerking points and the recent history, the wave force may be reduced to
\begin{equation}\label{eq:minimal}
    \bm{F}_W \sim \bm{F}_L + \bm{F}_N.
\end{equation}
The local force, $\bm{F}_L$ \eqref{eq:local}, captures the effect of the \markblue{droplet's} recent history, whereas the non-local force, $\bm{F}_N$ \eqref{eq:jp1}, captures the lack of wave cancellation at a discrete set of points in the \markblue{droplet's} past. Note that the identification and quantification of the jerking-point force $\bm{F}_N$ has remained elusive because it appears only as a beyond-all-orders  correction \cite{beyond_all_orders_2012} to the local wave force $\bm{F}_L$ \markblue{[i.e. it arises only after all orders in a local Taylor series expansion of the wave kernel at $s=0$, owing to the exponential suppression by the memory factor $\text{exp}(-(1-\Gamma)s_j)$]}. Nevertheless, it plays an essential role in driving the oscillatory dynamics of \markblue{walking droplets}.
Going forward, we refer to \eqref{eq:eom3} as the `full' [pilot-wave] model when using \eqref{eq:eom4} to evaluate $\bm{F}_W$, and the `minimal' model when using \eqref{eq:minimal} [since this omits all parts of the \markblue{droplet's} history where the waves excited cancel at the \markblue{droplet's} current location].

\section{Unified Mechanisms}

\subsection{In-line Speed Oscillations}

\begin{figure}[!t]
    \centering
    \includegraphics[width=\linewidth]{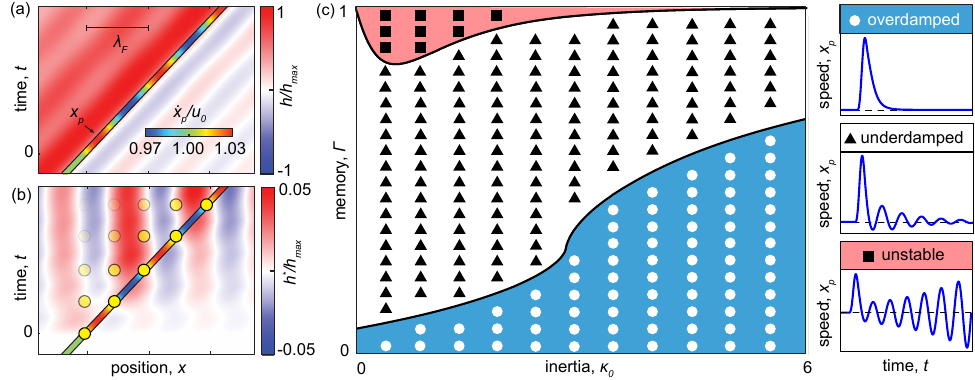}
    \caption{(a) Space-time diagram illustrating a one-dimensional \markblue{droplet} trajectory, $x_p(t)$ [colored by the local speed], with its wave field, $h(t,x)$. (b) An initial jerking point [at $t=0$] generates an oscillatory disruption of the wave field $h^*(t,x)$ that induces speed oscillations. These oscillations produce additional jerking points which once again give rise to speed oscillations, culminating in a self-reinforcing cascade of jerking points. (c) Perturbations from the \markblue{droplet's} preferred speed may either be overdamped, underdamped, or unstable. Numerical simulations of the minimal wave model [circles, triangles, squares] yield dynamics [identified via dynamic mode decomposition \cite{dmd1,dmd2,dmd3}] that agree with the phase portrait [blue, white, red] predicted by a linear stability analysis of the full wave model \cite{durey_2020_speed_osc}.}
    \label{fig:3}
\end{figure}

A hydrodynamic analog of Friedel oscillations~\cite{saenz_2020} revealed that in-line speed oscillations~\cite{Wind-Willassen2013,bacot2019multistable} may give rise to quantum-like statistics in pilot-wave systems. In this analogy, a localized heterogeneity in the domain, such as a \markblue{submerged circular well,  induces a sudden acceleration of the droplet due to the effectively higher wave memory in the deeper region. After crossing the well, the droplet relaxes back to its preferred speed through slowly decaying, underdamped speed oscillations along the outgoing trajectory. Because the origin of these oscillations is spatially fixed by the well, they superpose coherently across repeated droplet–impurity interactions. As a result, an ensemble of such droplet trajectories yields a wave-like} modulation in the \markblue{droplet’s} position histogram, with a spatial period set by the Faraday wavelength $\lambda_F$ [Fig.\,\ref{fig:1}(c)]. Bacot \textit{et al.}~\cite{bacot2019multistable} suggested that such speed oscillations may arise from wave-memory effects, while Durey \textit{et al.}~\cite{durey_2020_speed_osc} later performed a stability analysis confirming that the constant-speed solution to the full model \eqref{eq:eom3} may be unstable to perturbations [such as those produced by changes in fluid depth \cite{saenz_2020}], leading to sustained speed oscillations with the same wavelength $\lambda_F$ as the underlying wave field \cite{durey_2020_speed_osc,durey_2020_hqft,durey_2021,valani_2018,valani_2021,dagan_2020}. 
While stability theory established a regime diagram for the growth or decay of in-line speed oscillations [Fig.\,\ref{fig:3}(c)], the underlying physical mechanism driving them remained unidentified. 
%Our identification of anomalous interference at jerking points and their associated non-local forces fill this gap.

Here, we \markblue{demonstrate} that in-line speed oscillations originate from anomalous interference at jerking points and the non-local forces they generate. \markblue{Such oscillations arise following a sudden change in the droplet's velocity, which in practical settings may occur for various reasons, including interactions with lateral boundaries \cite{Wind-Willassen2013}, variations in submerged topography \cite{saenz_2020}, or generic system noise \cite{durey_2020_speed_osc}. Since the specific origin of the initial perturbation is not essential, we trigger the oscillatory dynamics by applying a short-lived external force $\bm{F}$ to induce a rapid change in speed (i.e. the first jerking point)}, which produces a spatially oscillatory non-local force \eqref{eq:jp1} with wavelength $\lambda_F$ [Fig.\,\ref{fig:3}(a-b)]. This force drives subsequent speed oscillations, each of which generates a new jerking point. The process thus repeats, yielding a cascade of jerking points with alternating signs [corresponding to transitions between acceleration and deceleration] every $\lambda_F/2$, which superpose coherently to produce the speed oscillations with wavelength $\lambda_F$ observed in experiments. The influence of each jerking point is governed by the memory of the wave field. At high memory, the oscillatory force persists for a longer time before decaying, allowing successive jerking points to interact more strongly and reinforce the oscillations. At low memory, the force dies out quickly, and the cascade is progressively dampened. Increasing the memory thus drives a transition from a regime of stable, decaying oscillations to one of unstable oscillations that grow in amplitude. We confirm this memory-dependent transition through direct numerical simulation of our minimal model, finding agreement with numerical simulations and a linear stability analysis of the full model \cite{durey_2020_speed_osc}[Fig.\,\ref{fig:3}(c)].

\begin{figure}[!t]
    \centering
    \includegraphics[width=\linewidth]{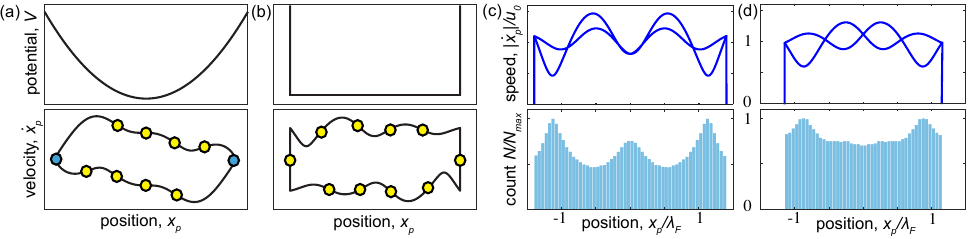}
    \caption{A \markblue{droplet} in a one-dimensional potential well exhibits speed oscillations driven by non-local forces anchored at the turning points. (a) In harmonic potentials, the turning points act as stationary points [blue] where the \markblue{droplet} momentarily pauses, triggering interior jerking points [yellow] that reinforce speed oscillations. (b) In square wells, turning points act as jerking points [yellow] due to the abrupt reversal of direction, producing wave-like statistics via a fundamentally different interference mechanism. Numerical simulations of our minimal model [$\kappa_0=2$, $\Gamma=0.8$] in square wells exhibit (c) constructive or (d) destructive interference, depending on the well width relative to $\lambda_F$.}
    \label{fig:4}
\end{figure}

\subsection{Quantum-like Statistics in Potential Wells}

Previous work~\cite{durey2018dynamics,montes_2019,montes_2021} has shown that when a walking droplet is confined to a one-dimensional harmonic potential \markblue{[$F = -kx_p$]}, its motion within the well is accompanied by oscillations in speed. These speed oscillations leave a coherent imprint on the position statistics of the \markblue{droplet}, producing wave-like modulations in the position histogram \markblue{[hence the terminology ``wave-like'', or ``quantum-like'', statistics]} whose spatial period matches the wavelength $\lambda_F$ of the underlying wave field. As the width of the well is increased [or, equivalently, as the restoring force is decreased], the wave-like position histogram alternates between pronounced oscillations and nearly uniform distributions, \markblue{akin to the discrete set of eigenmodes of a quantum particle in a one-dimensional well}. Large amplitude ripples emerge whenever the minima [maxima] in the \markblue{droplet's} speed coincide spatially with one another, which correspond to maxima [minima] in the position histogram. \markblue{This `resonance' condition occurs when a half-integer number [$n+1/2$] of wavelengths $\lambda_F$ fits within the well, while nearly uniform distributions arise when the well width equals an integer multiple  $n\lambda_F$.} Montes \textit{et~al.}~\cite{montes_2019,montes_2021} hypothesized that this resonance arises because the \markblue{droplet} spends more time near the turning points, the locations at the edges of the potential where the \markblue{droplet} reverses its direction. We note that this view is consistent with the analytical identification of stationary points at the turning points~\cite{blitstein_2024} [Fig.\,\ref{fig:4}(a)].

To demonstrate that wave-like statistics can emerge from a fundamentally different mechanism, we replace the smooth harmonic confinement with an infinite square well \markblue{[$F = 0$ for $|x_p|<L/2$ and $F = -\text{sign}(x_p)\infty$ for $|x_p|\geq L/2$, with $L$ the well width]}. In this setting, the \markblue{droplet} instantaneously reverses its velocity \markblue{without changing speed} at the boundaries [Fig.\,\ref{fig:4}(b)], so the turning points are not stationary points [where the \markblue{droplet} slows down and speeds up smoothly] but jerking points, where its direction of motion changes abruptly. These jerking points generate spatially oscillatory non-local forces with wavelength $\lambda_F$, directly imprinting the \markblue{droplet's} spatial statistics with a wave-like pattern [Fig.\,\ref{fig:4}(c-d)]. Each speed oscillation within the well triggers additional jerking points in the interior, creating a self-reinforcing cascade that shapes the oscillatory position histogram. Thus, the resulting wave-like structure in square wells does not stem from conventional constructive interference, as in harmonic wells, but from anomalous interference, in which jerking points act as persistent sources of oscillatory forces. Simulations of the full model agree with those of the minimal model shown in Fig.~\ref{fig:4}, confirming this picture.

\begin{figure}[!t]
    \centering
    \includegraphics[width=\linewidth]{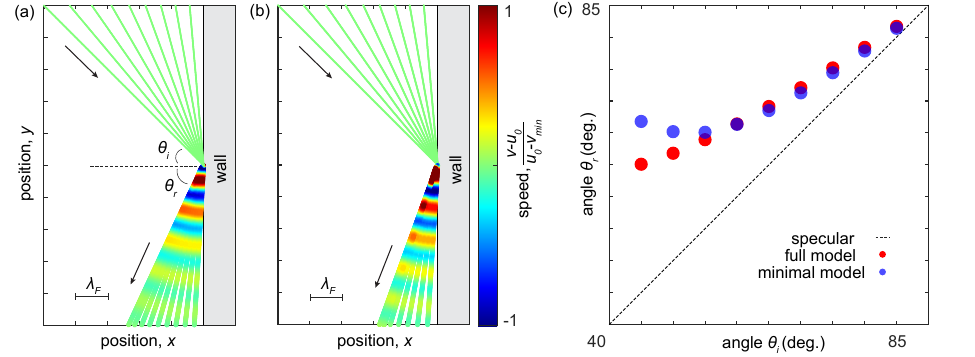}
    \caption{Non-specular reflection off a perfectly reflecting wall for the (a) full and (b) minimal  models [$\kappa_0=2,\Gamma=0.8$]. Besides a small phase shift, the speed oscillations induced by the jerking point at the point of reflection agree between both models. (c) Quantitative agreement is also found for the reflection angle, $\theta_r$, as a function of the incident angle, $\theta_i$, away from $45^{\circ}$ where our asymptotic approximations begin to break down \cite{SM}.}
    \label{fig:5}
\end{figure}

\subsection{Non-specular Reflection}

To illustrate that, in two or more spatial dimensions, anomalous interference can emerge from the geometry of the trajectory [owing to the dependence of the wave-surfing frequency \eqref{eq:frequency} on trajectory curvature and orientation, Fig.\,\ref{fig:2}] and produce effects not obviously tied to speed modulations, we consider the dynamics of a \markblue{walking droplet} encountering a perfectly reflecting wall \markblue{[$\bm{F}=\bm{0}$ for $\bm{x}_p\cdot\hat{\bm{x}} < 0$ and $\bm{F}=-\infty\hat{\bm{x}}$ for $\bm{x}_p\cdot\hat{\bm{x}} \geq 0$]}. At impact, the normal component of its velocity is instantaneously reversed, while the tangential component remains unchanged.

Even though the \markblue{droplet's} speed is unaltered, this instantaneous change in direction gives rise to a jerking point located at the reflection point [Fig.\,\ref{fig:5}], which generates a non-local force with two distinct effects. First, the jerking-point force induces in-line speed oscillations along the outgoing trajectory, despite there being no change in speed at the point of reflection. Second, because $\bm{\hat{d}}'(s)$ also changes discontinuously when the current position of the \markblue{droplet} is near the point of reflection, the jerking-point force can temporarily acquire a significant component normal to the trajectory, giving rise to non-specular reflection; the net reflection angle $\theta_r$ is larger than the incident angle $\theta_i$, even though $\theta_i=\theta_r$ at the point of reflection [Fig.\,\ref{fig:5}(a-b)]. Though formally outside the scope of our asymptotic approximations since $d(s_j) \lesssim 1$ when the normal force emerges, we find good agreement in simulations when $\bm{\hat{d}}(s)$ is treated as piecewise constant at the reflection point, in the same manner as in \eqref{eq:dp_ideal} [see \cite{SM}].
Numerical simulations of both the full and minimal models confirm these predictions, showing quantitative agreement for large $\theta_i$ [Fig.\,\ref{fig:5}(c)]. At smaller incident angles, additional differences arise due the influence of nearby stationary points when $\theta_i \lesssim 45^\circ$. \markblue{In this regime, the droplet turns through more than $90^\circ$, leading to a geometric stationary point with $\cos\theta(s_j)=0$ in \eqref{eq:frequency}, where the displacement vector from the past source to the current droplet location is perpendicular to the velocity at the source [see Fig.~\ref{fig:2}(b)]. Capturing this} regime properly would require a uniform asymptotic approximation between both the stationary and jerking-point contributions \cite{SM}. Interestingly, similar non-specular reflections have been observed experimentally for droplets reflecting off submerged walls~\cite{nonspecular}, suggesting that jerking points may play an important role in wave–boundary interactions in \markblue{walking droplets} \cite{harris_2013_b,abraham2024classical}.

\section{Conclusion}

Drawing from walking droplets, a prototypical model for wave-dressed active particles, we have uncovered a previously unrecognized form of wave interference that governs a variety of dynamical behaviors in these systems. Rooted in generic wave superposition, this anomalous interference arises when the frequency with which the \markblue{droplet} surfs its own wave field changes abruptly at discrete spatial locations, termed jerking points, producing a wave-mediated non-local force \eqref{eq:jp1} that oscillates with distance. This jerking-point force seeds a cascade of alternating
forces along the \markblue{droplet's} past trajectory,
with sign changes every $\lambda_F/2$, resulting in oscillatory dynamics with spatial period matching the Faraday wavelength $\lambda_F$ of the underlying wave field. Moreover, in two or more dimensions, the anomalous interference is sensitive to path geometry, thereby greatly broadening the range of dynamics shaped by the jerking-point force.

By deriving an asymptotic approximation of the jerking-point force from first principles and combining it with the local force from the most recent waves, which drives an overdamped relaxation to a preferred speed, we developed a minimal model that captures the essential ingredients underlying a range of wave-particle phenomena \markblue{that have largely been treated independently}. In one dimension, the jerking-point cascade accounts for the sustained in-line speed oscillations \cite{Wind-Willassen2013,bacot2019multistable,durey_2020_speed_osc} that serve as a precursor for the wave-like particle statistics observed in hydrodynamic analogs of Friedel oscillations \cite{saenz_2020}, for quantum-like position statistics in confining potentials \cite{durey2018dynamics,montes_2019,montes_2021}, and for spontaneous transitions into stochastic regimes at high memory \cite{hubert2019tunable,bacot2019multistable,hubert2022overload,durey_2020_speed_osc,durey_2020_hqft,durey_2021,valani_2018,valani_2021,dagan_2020}. In two or more dimensions, the same anomalous interference mechanism explains how geometry alone, without any change in speed, can trigger jerking points, leading to effects such as the non-specular reflection of \markblue{droplets} from walls \cite{nonspecular}, which one would not typically associate with the same force responsible for speed oscillations.

\markblue{Our findings also raise new questions for future work. Both Friedel oscillations \cite{saenz_2020} and non-specular reflections \cite{nonspecular} have thus far been observed in settings where the droplet interacts with submerged topography or lateral boundaries. Since we show that jerking points alone can generate these effects, it remains to be determined to what extent the observed speed oscillations in such experiments are driven by jerking points, as opposed to wave-mediated interactions with topography or boundaries.
The geometric sensitivity of jerking points further} raises the question of whether they might also underlie the sudden turns in the lemniscate and trefoil orbits that give rise to double quantization in a two-dimensional harmonic potential \cite{perrard_2014, kurianski_2017, durey_2017}.

\markblue{More broadly, several extensions could generalize the explicit form of the jerking-point force derived here by relaxing its underlying assumptions. In particular, incorporating the far-field spatial decay of the wave field \cite{molavcek_2013,milewski2015faraday,damiano2016surface} is expected to modify both the amplitude and phase shift of the jerking-point force, and may be important for accurately extending our framework to the interaction between multiple walking droplets. 
Similarly, relaxing the resonant walking assumption [i.e. allowing the impact phase $\mathcal{S}$ to vary with time \cite{molavcek_2013,galeano2018ratcheting}] may introduce additional effects. Several works have shown that the impact phase may depend on memory and the local wave height \cite{OrbitingPairs2017,arbelaiz2018promenading,couchman2019bouncing,couchman2020free}, or exhibit stochastic transitions \cite{harris_2013_b,Saenz2018,primkulov2025nonresonant}. Such non-resonant effects could introduce additional jerking points due to sudden changes in the amplitude, rather than the frequency, of the wave kernel. It seems plausible that such effects could be incorporated within our current framework by adding the endpoint contributions associated with these sudden changes in amplitude [Appendix \ref{sec:AppendixA}]. A systematic exploration of these extensions offers a path towards a more comprehensive theory.}
%It will also be of interest to explore how variable bottom topography may influence the jerking point cascade, particularly in settings relevant to hydrodynamic analogs of Friedel oscillations \cite{saenz_2020} and wave-like statistics in two-dimensional corrals \cite{harris_2013_b,cristea_2018,durey_2020_corral}. A systematic exploration of these extensions lies beyond the scope of the present work, but offers a path towards a more comprehensive theory.}

Although our analysis is rooted in the walking-droplet system \cite{couder_2005,bush_2015,bush2024perspectives}, the ingredients that give rise to the anomalous interference identified here, specifically wave interference and memory [i.e. non-zero wave decay time],
may be shared by a growing class of active pilot-wave systems. 
%In biology, for instance, honeybees trapped on the surface of a pond will flap their wings to generate hydrodynamic thrust through capillary waves, propelling themselves to survive \cite{honeybees}. On an entirely different scale, a person may jump periodically on a canoe to glide across a body of water, surfing the gravity waves generated by the bouncing \cite{gunwale,benham2024wave}. Inspired by these, scientists are currently designing aquatic robots that self-propel and interact with their environment via surface waves \cite{longuet1977mean,surfer,tarr_2024}. Beyond air-water interfaces, submersibles subject to pressure changes may navigate between layers of stratified salt water, riding along self-generated internal waves \cite{ludion}. An oscillating bubble may also self-propel by interacting with its radiated acoustic field \cite{acoustic,acoustic2}. In each of these cases, the active particles function as self-propelled wave sources, and abrupt changes in the relationship between their motion and the phase of its wave field may trigger non-local forces analogous to the jerking-point force derived here. Extending our framework to the specific features of each of these systems thus offers a promising direction for future research, and more broadly, invites consideration of anomalous interference as a potential organizing principle in wave-mediated active matter.
Examples span a wide range of scales, from insects generating capillary waves to propel themselves at fluid interfaces \cite{honeybees}, to macroscopic wave-surfing via periodic forcing \cite{gunwale,benham2024wave}, and to engineered systems such as wave-driven aquatic robots \cite{longuet1977mean,surfer,tarr_2024}. Related mechanisms may also arise in stratified fluids through internal waves \cite{ludion}, or in acoustically driven systems where particles interact with their own radiated fields \cite{acoustic,acoustic2}.
In all such cases, active particles act as self-propelled wave sources, and abrupt changes in the coupling between their motion and the phase of its wave field may generate non-local forces analogous to the jerking-point force identified here. Extending the present framework to these systems offers a promising direction for future work and, more broadly, invites consideration of anomalous interference as a potential organizing principle in wave-mediated active matter.
\vskip6pt

\enlargethispage{20pt}

\ack{This work is supported by the U.S. National Science Foundation through NSF CAREER Award CBET-2144180, the U.S. Office of Naval Research through Grant No N000142612109, and the Alfred P. Sloan Foundation through a Sloan Research Fellowship. We thank A. J. Abraham for Fig.\,\ref{fig:1}(a).}

\section{Appendices}

\subsection{Appendix A. Jerking Points\label{sec:AppendixA}}

We present the derivation of the jerking point force in full detail here. To begin, consider an idealized trajectory where the change in frequency at each jerking point $s_j$ is both instantaneous and between two constant values, the most recent, $d'_{j,a}$, and the following, $d'_{j,b}$, as in Fig.~\ref{fig:1}(g). We assume each jerking point is sufficiently far from the droplet's current location [$d(s_j) \gg 1$] so that we may use the large argument asymptotic expansion $2J_1(d)\sim b(d)\cos(d+\phi)$. We further assume the droplet speed is slowly varying outside of the time intervals $\Delta t_j$ [described in the Jerking Points section] during which the frequency rapidly changes near each jerking point [consistent with the same limit used to derive the local wave force] and $O(1)$ [since the droplet speed spends most of its time near the preferred value $u_0$]. 
Under these assumptions, the surfing frequency may be written as $d'(s)=f(\epsilon s)$, where $f=O(1)$ and $0<\epsilon\ll 1$, and therefore the distance function becomes $d(s)=\frac{1}{\epsilon}D(\epsilon s)$ with $\frac{dD(\xi)}{d\xi}=f$.
Note that a similar argument applies to the displacement vector $\bm{d}(s)=\frac{1}{\epsilon}\bm{D}(\epsilon s)$, so that the unit vector in that direction is likewise a slowly varying function $\hat{\bm{d}}(s)=\bm{D}(\epsilon s)/D(\epsilon s)$. Similarly, since $b(d)= O(1/d^{1/2})$ as $d\rightarrow\infty$ \cite{AbramowitzBook,bender_orszag_1978}, we may write $b(d) = \epsilon^{1/2}\beta(D(\epsilon s))$ with $\beta=O(1)$ for $D\gg 1$.

The lack of wave cancellation at the droplet's current location is due to the mismatch in frequency at each jerking point, for which we split the wave force integral \eqref{eq:eom4} into sub-intervals $[s_j,s_{j+1}]$ [each bounded by consecutive jerking points, with the first being $[0,s_1]$] to accommodate the piecewise discontinuities in $d'(s)$ near each jerking point
\begin{equation}\label{eq:SM:dp_ideal}
    d'(s)=
    \begin{cases}
        d'_{j,a} & s \leq s_j \\
        d'_{j,b} & s > s_j
    \end{cases}.
\end{equation}
%
%\rublue{%
After splitting the integral into the sub-intervals $[s_j,s_{j+1}]$, we substitute the asymptotic expansion for $2J_1(d)$ to obtain \eqref{eq:FN}. Assuming that there are no stationary points [where $d'$ vanishes], we then use the method of integration by parts \cite{copson_1965,bender_orszag_1978} to calculate the dominant contributions [which arise from the end points of each interval]. As we show, contributions from the interior of each sub-interval are negligible to leading order, owing to rapid oscillations that induce near-complete cancellation. 

For convenience, we define $\bm{A}_0(\epsilon s)=b(d(s))\hat{\bm{d}}(s) = \epsilon^{1/2}\beta(D(\epsilon s))\bm{D}(\epsilon s)/D(\epsilon s)$ and $\frac{1}{\epsilon}\Phi(\epsilon s)=d(s)+i\alpha s=\frac{1}{\epsilon}\left[D(\epsilon s)+i\alpha \epsilon s\right]$, where we have made use of our previous assumptions to identify $\bm{A}_0(\epsilon s)$ and $\Phi(\epsilon s)$ as slowly varying [though we note this assumption breaks down precisely near jerking points].
Plugging this into \eqref{eq:FN} yields
\begin{equation}\label{eq:SM:ibp1}
    \bm{F}_W \sim \real\left\{e^{i\phi}\sum_j\int_{s_j}^{s_{j+1}}\bm{A}_0(\epsilon s)e^{\frac{i}{\epsilon}\Phi(\epsilon s)}\dd s\right\}.
\end{equation}
It will be convenient to perform a change of variables to the slow timescale $\tau=\epsilon s$, so that \eqref{eq:SM:ibp1} becomes
\begin{equation}
    \bm{F}_W \sim \real\left\{e^{i\phi}\sum_j\int_{\tau_j}^{\tau_{j+1}}\tilde{\bm{A}}_0(\tau)e^{\frac{i}{\epsilon}\Phi(\tau)}\dd \tau\right\},
\end{equation}
where $\tilde{\bm{A}}_0(\tau)=\frac{1}{\epsilon}\bm{A}_0(\tau)$.

We now integrate by parts once on each sub-interval, first multiplying and dividing by $\frac{i}{\epsilon}\Phi'(\tau)$, to obtain
\begin{equation}\label{eq:SM:ibp2}
    \begin{aligned}
    \bm{F}_W &\sim \real\left\{e^{i\phi}\sum_j\int_{\tau_j}^{\tau_{j+1}}\frac{\tilde{\bm{A}}_0(\tau)}{\frac{i}{\epsilon}\Phi'(\tau)}\left(e^{\frac{i}{\epsilon}\Phi(\tau)}\right)'\dd \tau\right\} \\&
    = \real\left\{e^{i\phi}\sum_j\left[\frac{\tilde{\bm{A}}_0(\tau)e^{\frac{i}{\epsilon}\Phi(\tau)}}{\frac{i}{\epsilon}\Phi'(\tau)}\eval_{\tau_j}^{\tau_{j+1}}-\epsilon\int_{\tau_j}^{\tau_{j+1}}\tilde{\bm{A}}_1(\tau)e^{\frac{i}{\epsilon}\Phi(\tau)}\dd \tau\right]\right\},
    \end{aligned}
\end{equation}
where we define $\tilde{\bm{A}}_1(s)=(\tilde{\bm{A}}_0(\tau)/i\Phi'(\tau))'$ which is $O(1)$ within the interior of each sub-interval, and have assumed no stationary points where $\Phi'=0$; the contributions from stationary points [if any] need to be added separately, and were computed in \cite{blitstein_2024}. The new integral involving $\tilde{\bm{A}}_1(\tau)$ contributes an $O(\epsilon^2)$ term, as may be seen by integrating by parts in the same way once more,
\begin{equation}\label{eq:SM:ibp3}
    \epsilon\int_{\tau_j}^{\tau_{j+1}}\frac{\tilde{\bm{A}}_1(\tau)}{\frac{i}{\epsilon}\Phi'(\tau)}\left(e^{\frac{i}{\epsilon}\Phi(\tau)}\right)'\dd \tau = \epsilon\left[\frac{\tilde{\bm{A}}_1(\tau)e^{\frac{i}{\epsilon}\Phi(\tau)}}{\frac{i}{\epsilon}\Phi'(\tau)}\eval_{\tau_j}^{\tau_{j+1}}-\epsilon\int_{\tau_j}^{\tau_{j+1}}\tilde{\bm{A}}_2(\tau)e^{\frac{i}{\epsilon}\Phi(\tau)}\dd \tau\right] = O\left(\epsilon^2\right),
\end{equation}
where $\tilde{\bm{A}}_2(\tau)=(\tilde{\bm{A}}_1(\tau)/i\Phi'(\tau))'$. Thus, the integral involving $\tilde{\bm{A}}_1(\tau)$ [which consists of the contribution from the interior and the beyond-leading order corrections to the endpoint contribution] is thus negligible [to leading order] relative to the $O(\epsilon)$ contribution arising from the boundary terms in \eqref{eq:SM:ibp2} [which capture the dominant endpoint contributions on each sub-interval]. 

After discarding the $O(\epsilon^2)$ integral and omitting the endpoint contribution near $s=0$ [which is computed separately in the local force section] we are left with the non-local contribution $\bm{F}_N$ from the jerking points. Grouping the right and left boundary terms associated with each jerking point $s_j$, from the intervals $[s_{j-1},s_j]$ and $[s_j,s_{j+1}]$ respectively, yields
\begin{equation}\label{eq:SM:jp1}
    \bm{F}_N \sim \real\left\{e^{i\phi}\sum_{j}\left[\left[\frac{\tilde{\bm{A}}_0(\tau)e^{\frac{i}{\epsilon}\Phi(\tau)}}{\frac{i}{\epsilon}\Phi'(\tau)}\right]\right]_{\tau_j}\right\} = \real\left\{e^{i\phi}\sum_{j}\left[\left[\frac{b(d(s))\hat{\bm{d}}(s)e^{id(s)-\alpha s}}{id'(s)-\alpha}\right]\right]_{s_j}\right\},
\end{equation}
where $[[f(s)]]_{s_j}=\lim_{\zeta\rightarrow 0^{+}}[f(s_j-\zeta)-f(s_j+\zeta)]$ denotes the jump discontinuity across $s_j$, and we have substituted back $\tilde{\bm{A}}_0(\tau)=\frac{1}{\epsilon}b(d(s))\hat{\bm{d}}(s)$ and $\frac{1}{\epsilon}\Phi(\tau)=d(s)+i\alpha s$. Since only $d'(s)$ is discontinuous at each $s_j$ in our instantaneous change approximation \eqref{eq:SM:dp_ideal}, \eqref{eq:SM:jp1} thus reduces to
\begin{equation}\label{eq:SM:jp2}
    \bm{F}_N \sim \real\left\{e^{i\phi}\sum_{j}\left[\frac{1}{id'_{j,a}-\alpha}-\frac{1}{id'_{j,b}-\alpha}\right]b(d(s_j))\hat{\bm{d}}(s_j)e^{id(s_j)-\alpha s_j}\right\},
\end{equation}
which may be written more compactly using \eqref{eq:jp2} as
\begin{equation}\label{eq:SM:jp3}
     \bm{F}_N \sim \sum_j D_j\Big[b(d(s_j))\cos(d(s_j)+\theta_j+\phi)\Big]e^{-\alpha s_j}\,\hat{\bm{d}}_j.
\end{equation}

As it stands, \eqref{eq:SM:jp3} offers one approximation to the jerking point force. However, this formula incorrectly diverges to infinity when the jerking point is near the droplet, where $d(s_j)$ is small, because in the large-argument asymptotic form of the wave field, $2J_1(d)\sim b(d)\cos(d+\phi)$, involves an amplitude $b(d)$ that typically becomes singular at the origin $d=0$. To extend the validity of our force approximation to smaller values of $d(s_j)$, we recast \eqref{eq:SM:jp3} in terms of $2J_1(d)$, rather than its asymptotic approximation for large $d$. To do so, we exploit the fact that, since $b(d)$ is slowly varying for $d\gg 1$ [$(\text{d}b/\text{d}d)/b\ll 1$], it follows that $b(d)\sim b(d+\theta)$ as $d\to\infty$ for any constant phase shift $\theta$. This allows us to conclude that $b(d(s_j))\cos(d(s_j)+\theta_j+\phi)\sim 2J_1(d(s_j)+\theta_j)$ for large $d(s_j)$, which we substitute into \eqref{eq:SM:jp3} to obtain \eqref{eq:jp1}. Since $2J_1(d)$ is regular at the origin, the resulting expression provides an asymptotic approximation of $\bm{F}_N$ that remains well-behaved even when  $d(s_j)$ is small, in contrast to \eqref{eq:SM:jp3}. For non-idealized trajectories in which the frequency varies continuously rather than instantaneously, we refer to the corresponding discussion in the Non-Idealized Trajectory section.

\subsection{Appendix B. Limitations of Jerking Points for Non-Specular Reflection \label{sec:AppendixB}}

Occasionally, when the droplet is close to a jerking point, $\hat{\bm{d}}(s)$ may rapidly change direction, as is the case immediately following reflection off a wall in Fig.~\ref{fig:5}. Since $\hat{\bm{d}}(s) = (\cos\Theta(s),\sin\Theta(s))$, where $\Theta(s)$ is the polar angle of the two dimensional unit vector $\hat{\bm{d}}(s)$, a more precise calculation of $\bm{F}_N$ would account for this additional phase $\Theta$ in the exponential in \eqref{eq:SM:jp1}, which itself could shift the location of existing jerking points or even introduce new ones. However, since \eqref{eq:SM:jp1} is already able to account for an additional discontinuous change in $\hat{\bm{d}}(s)$ near jerking points, we instead
approximate $\Theta(s)$ as piecewise discontinuous by averaging $\Theta(s)$ over a suitable interval on either side of $s_j$ to get $\Theta_{j,a}$ and $\Theta_{j,b}$, which we then treat as constant on that side of the jerking point.
Defining $\hat{\bm{d}}_{j,X} = (\cos\Theta_{j,X},\sin\Theta_{j,X})$ and $D_{j,X}e^{i\theta_{j,X}} = 1/(id'_{j,X}-\alpha)$ for $X=a,b$, we find
\begin{equation}\label{eq:SM:reflection}
\begin{split}
    \bm{F}_N \sim \sum_j \Big[D_{j,a}2J_1(d_{j}+\theta_{j,a})\,\hat{\bm{d}}_{j,a}
    - D_{j,b}2J_1(d_{j}+\theta_{j,b})\,\hat{\bm{d}}_{j,b}\Big]e^{-\alpha s_j}.
\end{split}
\end{equation}
Numerical simulations indicate good agreement with the full wave model in Fig.~\ref{fig:5} when we average $\Theta(s)$ over the time it takes the droplet to travel a distance $\lambda/4$ on either side of the jerking point.

\subsection{Appendix C. Explicit Formulas for Numerical Simulations \label{sec:AppendixC}}

We state explicitly the form of all the wave forces that enter into the full and minimal wave models. The full wave force on the droplet is
\begin{equation}\label{eq:SM:FW}
    \bm{F}_W = 2\int_{0}^{\infty}J_1(d(s))\hat{\bm{d}}(s)\,e^{-(1-\Gamma)s}\dd{s}.
\end{equation}
The local wave force is
\begin{equation}\label{eq:SM:FL}
    \bm{F}_L \sim \frac{2}{|\dot{\bm{x}}_p|^2}\left[1-\frac{1}{\sqrt{1+(|\dot{\bm{x}}_p|/(1-\Gamma))^2}}\right]\dot{\bm{x}}_p.
\end{equation}
The non-local wave force is a sum over all jerking and stationary points
\begin{equation}\label{eq:SM:FN}
    \bm{F}_N \sim 2\sum_j D_j J_1(d(s_j)+\theta_j)e^{-(1-\Gamma) s_j}\,\hat{\bm{d}}_j
\end{equation}
with
\begin{equation}\label{eq:SM:Djp}
    D_j e^{i\theta_j} = \frac{1}{id'_{j,a}-(1-\Gamma)} - \frac{1}{id'_{j,b}-(1-\Gamma)}
\end{equation}
if $s_j$ is a jerking point, or
\begin{equation}\label{eq:SM:Dsp}
    D_j e^{i\theta_j} = \sqrt{\frac{2\pi}{|d''_j|}}e^{i\,\text{sign}(d''_j)\left(\frac{(1-\Gamma)^2}{2|d''_j|} + \frac{\pi}{4}\right)}
\end{equation}
if $s_j$ is a stationary point. The contributions from stationary points were previously derived in \cite{blitstein_2024}, though with a slight difference: here we evaluate the direction vector $\hat{\bm{d}}(s)$ as a constant evaluated at $s = s_j$, consistent with the treatment used for jerking points. In contrast, \cite{blitstein_2024} evaluates $\hat{\bm{d}}(s)$ at the nearby saddle point in the complex plane associated with the stationary phase approximation. The main difference between the approaches is that the one used here neglects the relatively small component of the force perpendicular to the displacement vector $\hat{\bm{d}}(s_j)$, which arises from the slight asymmetry in the exponential memory kernel that favors more recent contributions $(s<s_j)$ over older ones $(s>s_j)$. As discussed in \cite{blitstein_2024}, this correction  has only a minor influence on the overall dynamics, since larger forces [either from stationary points if $s_j$ is a jerking point, or the local force if $s_j$ is a stationary point] already act along that direction. We therefore neglect it here, though it may prove relevant for future analyses of stability.

%%%%%%%%%% Insert bibliography here %%%%%%%%%%%%%%

\vskip2pc

\bibliographystyle{RS} %%%% .BST file

\bibliography{bibliography} %%%%% .Bib file

\end{document}

%%%% Article title to be placed here
\title{Supplemental Material for: Anomalous interference drives oscillatory dynamics in wave-dressed active particles}

\author{%%%% Author details
Austin M. Blitstein$^{1}$, Rodolfo R. Rosales$^{2}$, Pedro J. S\'{a}enz$^{1}$}

%%%%%%%%% Insert author address here
\address{$^{1}$Department of Mathematics, University of North Carolina, Chapel Hill, North Carolina, 27599, USA\\
$^{2}$Department of Mathematics, Massachusetts Institute of Technology, Cambridge, Massachusetts, 02139, USA}

%%%% Subject entries to be placed here %%%%
\subject{applied mathematics}

%%%% Keyword entries to be placed here %%%%
\keywords{active particles, wave interference, path memory, walking droplets}

%%%% Insert corresponding author and its email address}
\corres{Pedro J. S\'{a}enz\\
\email{saenz@unc.edu}}

%%%%%%%%%%%%%%%%%%%%%%%%%%%

%\rsbreak

\maketitle

%%%%%%%%%% Insert the texts which can accomdate on firstpage in the tag "fmtext" %%%%%

\section{Archetypical Framework}

We demonstrate that the net wave force acting on a wave-dressed active
particle takes an archetypical form analogous to the one we use for walking droplets in the main text under the following conditions: (1) The particle may be idealized as a point in space that undergoes fast intrinsic oscillations [with characteristic time scale $T_o$] that excite waves in an ambient
field\,\footnote{By oscillating we mean a repetitive function, characterized by a typical time scale $T_o$, for which the notion of an average value is defined. For example: periodic [then $T_o$ is the period] or quasi-periodic [then $T_o$ is the longest period].\label{footnoteTo}}.
% rrr We had $T_{\rm osc}$ in the footnote and $T_o$ in the text. Fixed it.
%     I added a label because we cite this footnote later.
This wave field then produces a force on particles by some mechanism. (2) The particle translates slowly, with typical speed $v$. (3) The waves are small amplitude, dissipative, and their equations of motion are invariant under translation and rotation. (4) There exists a mechanism through which a single dominant wavelength $\lambda$ emerges, with associated wavenumber $k=2\pi/\lambda$. We elaborate later on how this condition may be relaxed
% rrr. I think an "on" was missing before how.
to include a finite number of dominant wavelengths. (5) The particle's intrinsic oscillations are in resonance with the selected wave number $k$. (6) The decay rate for wave modes which persist on timescales much longer than the timescales of the oscillations present in the system can be approximated as
exponential\,\footnote{When the coefficients of the equations governing the wave field are periodic in time, Floquet theory \cite{GrimshawODE,InceODE} can be used to show that the decay rate for each wave mode is exponential, as is the case for the walking droplet system. In more general situations, such as walking droplets with a multi-frequency forcing,  if the coefficients are oscillatory and ``steady'' [the controlling parameters do not change], it is reasonable to assume that the decay rate [for long lived modes] produced by the dissipation is also exponential. This assumption is based on these modes ``seeing'' only average quantities, which is reasonable for the modes that decay much slower than the oscillations time scales for both the coefficients and solution.}.
(7) To leading order [linear approximation], the average force field generated by a particle is nonzero.

Before showing how these conditions lead to Eq.~2.1-2.3 in the main text, we make some clarifying remarks. The wave amplitudes are small, which allows us to linearize the equations governing their motion. Owing to translation invariance, the waves can be written as a linear combination of Fourier eigenmodes: elementary solutions whose space dependence has the form $e^{i\bm{q}\cdot\bm{x}}$, where  $\bm{q}$ is the wavenumber. Note that the time dependence of these modes need not be exponential \footnote{The equations governing the waves can have coefficients that are oscillatory in time. Hence, the time dependence of the Fourier eigenmodes need not be exponential [i.e. of the form $e^{i\omega\,t-\rho t}$] as happens when the coefficients are constant. However, we assume that this time dependence is oscillatory, with a superimposed exponential decay.}.
%
% rrr We had "it" is oscilatory. Better be clear. Added exponential to decay.
%
Rotation invariance also implies that the time dependence of the Fourier eigenmodes is a function of $|\bm{q}|$ only. The existence of a wave selection mechanism means that Fourier eigenmodes with wavelengths away from $\lambda$ decay much faster, say over time
scales equal or less than $T_s$, with $T_s \ll T_M$, where $T_M$ is the time scale over which waves with wavelengths in a narrow band centered on $\lambda$ decay. Since $T_p = \lambda/v$ is the timescale associated with the particle translation,
slow translation means $T_s\/,\,T_o \ll T_p$, and fast oscillations means $T_o \ll T_M$\,\footnote{Note that $T_o \ll T_M$
is also needed for the particle to be able to excite any waves.}. Owing to the particle's intrinsic oscillations being in resonance with the selected wavenumber $k$, the time scale of oscillations associated with the wavenumber $k$ is $T_o$. Further, since $T_o \ll T_M$, the Fourier eigenmodes with the selected wavelength decay exponentially [condition 6]. Thus, the associated Fourier eigenmodes take the form $\bm{b}(t)\,e^{i\,\bm{q}\cdot\bm{x} - t/T_M}$, where $\bm{b}(t)$ is
oscillatory with typical time scale $T_o$, $|\bm{q}|=k$, and $\bm{b}$ is the same for all $\bm{q}$. Finally, because the waves are small amplitude, we can approximate the mechanism through which the waves generate a force as linear --- hence
the force field inherits the Fourier eigenmode structure of the waves.

Consider now the wave field, $\bm{w}(\bm{x}, t)$, generated by a slowly translating particle at $\bm{x} = \bm{x}_p(t)$, starting at some time $t_0$. Duhamel's principle \cite{GuentherPDE} then allows us to write $\bm{w}(\bm{x}, t)$ in terms of the solution to initial value problems: $\bm{w} = \bm{w}_0+\int_{t_0}^t \bm{W}(|\bm{\dis}|\/,\,t\/,\,\tau)\,\dd\tau$,
where $\bm{w}_0$ is the wave generated by the wave initial conditions at $t = t_0$, $\bm{W}$ is the solution [to the homogeneous wave equation] with initial conditions at $t = \tau$ given by the forcing from the particle at $t = \tau$,
and we have implemented rotation and translation invariance by displaying $\bm{W}$ as a function of $|\bm{\dis}| = |\bm{x}-\bm{x}_p(\tau)|$ only. Owing to wave selection, we can eliminate the Fourier eigenmodes from $\bm{W}$ with $|\bm{q}| \neq k$. This simplification is justified for a slow translating particle since the time it takes the particle to translate away from its current position is $t-\tau=O(T_p)\gg T_s$, and the assumption that $|\bm{\dis}|$ is not too large.
%
% rrr We had "because". It is an assumption, we are not looking too far.
% Only when you have multiple drops "look far" becomes important.
% Note that $|\bm{\dis}|$ not too large is why we can use Bessel.
% Otherwise decay with distance in the kernel is needed.
%
We can thus write
\begin{equation}\label{eq:SM:SingleIV}
 \bm{W} \sim a(\tau)\,\bm{b}(t)
 \int e^{i\,\bm{q}\cdot\bm{\dis} - (t-\tau)/T_M}\,
 \delta\left(|\bm{q}|-k\right) \dd^n\bm{q},
\end{equation}
where $a(\tau)$ is determined by the ``phase'' of the intrinsic particle oscillations at time $t=\tau$ [and it is independent of the wavenumber because of rotation invariance], and $n$ is the number of spatial
dimensions\,\footnote{The exact $\bm{W}$ in Eq.~\ref{eq:SM:SingleIV} involves an integral over all the Fourier eigenmodes, each with a weight $\tilde{a}(\tau, |\bm{q}|)\,e^{-i \bm{q}\cdot\bm{x}_p(\tau)}$ provided by the initial conditions [$\tilde{a}$ depends on $|\bm{q}|$ only because of rotation invariance]. Then we use the wave selection property, and keep only the modes with $|\bm{q}|=k$. This approximation is a bit rough, because generally there is a narrow band of modes near $|\bm{q}|=k$ with decay rates not too different from $1/T_M$. Lumping them all into $|\bm{q}|=k$ introduces errors that can become significant for $|\bm{\dis}|$ large; e.g.: $\bm{W}$ should decay exponentially as $|\bm{\dis}| \to \infty$ [because of the dissipation], which Eq.~\ref{eq:SM:SingleIV} misses. Incorporating corrections for these sort of errors is difficult, as the literature in walking drops demonstrates; e.g. see \cite{MolavcekThesis2013,TopoMeasures2016,OrbitingPairs2017,%
BouncingPhaseVar2019,TowardsGeneralizedPilot2018,CouchmanThesis2020}. Finally, we should be writing $\bm{x} = \bm{x}_p(t) + \delta\bm{x}_p(t)$ for the particle position, where $\delta\bm{x}_p(t)$ accounts for the fluctuations caused by the oscillations in the force by the waves. However, since these oscillations are fast, the velocity fluctuations are small [relative to $v$], and the position fluctuations are even smaller [relative to $\lambda$]. Thus we neglect
$\delta\bm{x}_p(t)$ for the leading order approximation in Eq.~\ref{eq:SM:SingleIV}.}.

Since $a(\tau)$ must be oscillatory \footnote{See footnote~\ref{footnoteTo}.}
% rrr. Here we had: See footnote 6. uh?
%
with time constant $T_o \ll T_M$, and the particle translates sufficiently slow relative to the oscillations, $T_o \ll T_p$, we can replace $a(\tau)$ in the integral $\int_{t_0}^t \bm{W} \dd\tau$ above by its average $\overline{a}$\;\footnote{Replacing $a$ by $\overline{a}$ in
Eq.~\ref{eq:SM:TheWave} is justified because in Eq.~\ref{eq:SM:SingleIV} the only fast $\tau$ dependence enters through
$a(\tau)$. The others, via $\bm{x}(\tau)$ and $\tau/T_M$ in the exponent, are slow. The calculation behind this is simple: Let $0 < \epsilon \ll 1$, $g$ oscillatory, and $f$ integrable. Then $\int_{t_0}^t g(s/\epsilon)f(s)\,\dd\/s
= \overline{g}\int_{t_0}^t f(s)\,\dd\/s + O(\epsilon)$}, which leads to
\begin{equation}\label{eq:SM:TheWave}
 \bm{w} \sim \bm{w}_0 + \bm{b}(t)\,\int_{t_0}^t
          A_w\left(|\bm{\dis}|\right)\,e^{-(t-\tau)/T_M}\,\dd\tau,
\end{equation}
where $A_w\left(|\bm{\dis}|\right) = \overline{a}\,\int \delta\left(|\bm{q}|-k\right)\,e^{i\,\bm{q}\cdot\bm{\dis}}\,
\dd^n\bm{q}$. Note that we have used the form of Duhamel's principle corresponding to a first order in time equation for the
waves; however there are analogs for any order [and higher order systems can always be reduced to first order by adding extra variables].

Every $\bm{W}$ in Eq.~\ref{eq:SM:SingleIV} generates a force on a test particle at any position $\bm{x}$ [condition 1] which must be radial [so as not to violate conservation of angular momentum in the system]. It follows that
the force field it generates must be derivable from a radial potential. Further, because we can approximate the force field as linear in the waves, the total force field is a linear superposition of the force fields generated by each individual $\bm{W}$, hence also potential. In fact, the force field will inherit the structure of Eq.~\ref{eq:SM:TheWave}. It
follows that we can write
\begin{equation}\label{eq:SM:ForceField}
 \bm{F} = \bm{F}_0 - \alpha(t) \grad \int_{t_0}^{t}
 A_U(|\bm{\dis}|)\,e^{-(t-\tau)/T_M}\,\dd\/\tau
\end{equation}
for some [scalar] functions $\alpha$ and $A_U$, where $\alpha$ is oscillatory with characteristic time $T_o$, and $\bm{F}_0$ is the force field generated by the wave initial conditions at time $t = t_0$. The average force field then follows by replacing $\alpha$ by $\bar{\alpha}$ [valid for $T_o \ll T_p,T_M$], yielding
\begin{equation}\label{eq:SM:AvForceField}
 \bm{F}_W = \bm{F}_0 - \grad \int_{t_0}^{t}
 A(|\bm{x}-\bm{x}_p(\tau)|)\,e^{-(t-\tau)/T_M}\,\dd\/\tau,
\end{equation}
which is the form of the net wave force used in the main text. Here $T_M$ is the ``effective force field'' decay time [or ``path memory''] time, and $A = \bar{\alpha}\,A_U$ is an effective potential associated with the force field generated by a stationary particle [in the context of walking droplets, this is the wave field associated with a drop that bounces in
place \cite{oza_2013,molavcek_2013}]. To see the connection between $A$ and the force field generated by a stationary particle, consider what happens when $\bm{x}_p$ is constant. Then $\lim_{t\to \infty} \bm{F}_W = -T_M\, \grad A(|\bm{x}-\bm{x}_p(\tau)|)$.

By assumption $\grad A$ does not vanish [condition 7]. This matters for the asymptotic homogenization process just described, which is based on a quasi-steady state approximation, and where $\bm{F}_W$ is the leading order force. If $\bm{F}_W$ were to vanish, higher order terms would have to be included in the model.

Intuitively, the uniform exponential damping, $e^{-(t-\tau)/T_M}$ in
Eq.~\ref{eq:SM:AvForceField}, arises [once the particle moves to a
different location] because all the Fourier eigenmodes of the wave
decay at the same averaged rate, corresponding to the single dominant
wavenumber $k$ [with time constant $T_M$], starting at the time the
particle moves to a new position.

We further demonstrate that $A$ is uniquely determined by the number of
spatial dimensions, $n$, up to a system-specific constant. As shown
earlier, because the waves are homogeneous and small amplitude, $A$ may
be written in terms of a linear superposition of Fourier modes. Let
$\tilde{A}(\bm{q})$ denote the Fourier transform of $A(k|\bm{\dis}|)$
with respect to $\bm{\dis}=\bm{x}-\bm{y}$. Since the waves are isotropic,
there can only be a dependence on the magnitude of the wavenumber,
$\tilde{A} = \tilde{A}(|\bm{q}|)$. Additionally, there is a single
dominant wavenumber $k$, so that $\tilde{A} = C\delta(|\bm{q}|-k)$ is
determined up to a constant $C$, where $\delta$ is the Dirac delta
function. It follows that
\begin{equation}\label{eq:AisBessel}
\begin{split}
A(k|\bm{\dis}|)
&= \frac{C}{(2\pi)^n} \int_{-\infty}^{\infty} e^{i\bm{q}\cdot\bm{\dis}}\,
   \delta(|\bm{q}|-k)\,\dd^n\bm{q} \\
&= \frac{Ck^{n-1}}{(2\pi)^n} \int_{S^{n-1}}
   e^{i\bm{k}\cdot\bm{\dis}}\,\dd\Omega^{n-1} \\
&= \frac{Ck^{n-1}}{(2\pi)^{n/2}} \frac{1}{(k|\bm{\dis}|)^{n/2-1}}\,
   J_{n/2-1}(k|\bm{\dis}|),
    \end{split}
\end{equation}
where $S^{n-1}$ is the $n-1$ dimensional unit sphere,
$\dd\Omega^{n-1}$ is its associated [hyper]surface area element, and
the last equality follows from one of the integral definitions of the
first-kind Bessel function $J_{\nu}$ of order $\nu$
\cite{AbramowitzBook,bender_orszag_1978}.
Eq.~\ref{eq:AisBessel} thus demonstrates the uniqueness of the
effective potential $A$ in $n$ spatial dimensions [up to a constant].
For the numerical simulations presented in the main text we set $n=2$,
resulting in $A(k|\bm{\dis}|) = C_0J_0(k|\bm{\dis}|)$, with $C_0$ a
system specific constant proportional to $C$. For reference, in the
cases $n=1$ and $n=3$, the effective potential becomes
$A(k|\bm{\dis}|) = C_0\cos(k|\bm{\dis}|)$ and
$A(k|\bm{\dis}|) = C_0\sin(k|\bm{d}|)/(k|\bm{\dis}|)$, respectively.

The assumption of a single dominant wavelength $\lambda$ may arise in a
number of different ways. For example, in walking
droplets \cite{molavcek_2013,oza_2013} the selection mechanism is caused
by periodic parametric forcing [Faraday waves \cite{KumarTuckerman1994}].
Via Floquet theory \cite{GrimshawODE} it can be shown that this selects
waves with one half the forcing frequency, which results in the decay
rate time [constant] for all wavenumbers with $|\bm{q}| \neq k_F$ being
much smaller than the decay rate time for wavenumbers with
$|\bm{q}| \approx k_F$ [here $k_F$ is the Faraday wavenumber].

Notice also that it is easy to extend the approach to situations where a
finite number of wavelengths are selected [instead of just one].
In such a case, one would treat each wavelength separately, and write the
total force as a linear combination of forces like that in Eq.~2.2 of the
main text. A situation like this would arise, for example, when the wave
selection is done via parametric forcing, but the selected wave frequency
corresponds to more than one wavelength.

In the absence of a wave selection mechanism, it may be possible to
achieve a situation where the waves are dominated by a single wavelength,
by forcing the field at a single frequency \cite{surfer,ludion}. In this case a point particle would likely not work.
One would have to use particles designed so that they predominantly
excite waves with a wavelength corresponding to the forcing frequency.
The reason that a point particle may not work in this scenario is that
such a particle would, generally, 
excite the full Fourier spectrum [because
the Fourier transform of a Dirac delta function has the same amplitude for all
Fourier modes]. However, it is still possible [using Duhamel's principle]
to write the force field as an integral over the past trajectory of the
particle, but the integrand [instead of taking the simple form in Eq.~2.2
of the main text] would involve the full Green's function for the
governing wave equation. An example of this can be found in
\cite{durey_2020_frontiers}.

Finally, we note that the exact details through which the wave generates
a force on a particle depends on the particular system under study. For
example, in walking drops the force has the form $-\sigma(t)\grad h$,
where $h$ is the surface elevation and $\sigma$ is an effective weight
that incorporates when the drop is in contact with the surface
\cite{molavcek_2013}. Here, the [oscillatory] time dependence of the
weight is crucial, as it makes the average [homogenized] force over a
bouncing period non-vanishing. Note that, in this case, $A$ is a
weighted average of $h$. Moreover, under certain conditions, $\sigma$
closely resembles a Dirac delta per bouncing period, in which case $A$
is nearly equivalent to $h$ evaluated at a particular moment of its time
period. This peculiarity motivates the term `strobed wave field' and
explains why the equation of motion based on this approximation is called
the `strobe model' \cite{oza_2013}.

\section{Uniform Asymptotic Approximation for Jerking and Stationary Points}

If the change in $d'(s)$ in the vicinity of a stationary point is sufficiently rapid [such as for non-specular reflections with incident angle less than $45^\circ$], the correct calculation for $D_je^{i\theta_j}$ is actually Eq.~(7.13) instead of Eq.~(7.14) in Appendix A in the main text. For intermediate rates of change, where neither approximation fully applies, a uniform asymptotic approximation combining both formulas is required. We note one possible approach here.
%\rured{\footnotesize\sf
%\\I am not sure reflection is a good example for this, as there the trajectory has an actual angle. A better example would be a trajectory that has a rapid, but smooth, bend. In the case of reflection, this would be the case when reflection occurs because of a steep potential wall --- not a hard wall as happens in the example of the main text.}

Let $s_j$ be the associated jerking point for a change in frequency. Instead of assuming an instantaneous change from $d'_{j,a}$ to $d'_{j,b}$, we assume a piece-wise linear change that agrees with first order Taylor expansions of $d'(s)$ about both extrema and the jerking point
\begin{equation}\label{eq:SM:dp_uniform}
    d'(s) \approx \begin{cases}
        d'_{j,a} & s_{j,a} \leq s \leq s_j-T_a \\
        d'_j + d''_j(s-s_j) & s_j-T_a \leq s \leq s_j+T_b \\
        d'_{j,b} & s_j+T_b \leq s \leq s_{j,b}
    \end{cases}
\end{equation}
where $T_{X} = |(d'_{j,X}-d'_j)/d''_j|$, with $X=a,b$, is the time it takes the first order Taylor approximation at the jerking point to intersect with the first order Taylor approximation at the extrema $s_X$. Upon splitting the integral into three parts associated to each section of Eq.~\ref{eq:SM:dp_uniform}, the exterior integrals may be calculated the same as in Eq.~(7.4) in Appendix A in the main text, and the interior integral may be calculated as a complex Fresnel integral $E(Z) = \int_0^Z e^{i\pi x^2/2}dx$, yielding
\begin{align}\label{eq:SM:D_uniform}
    D_j e^{i\theta_j} =& \frac{1}{id'_{j,a}-\alpha}e^{-i(d'_jT_a-\frac{1}{2}d''_jT_a^2)+\alpha T_a} \\\nonumber
    &- \frac{1}{id'_{j,b}-\alpha}e^{i(d'_jT_b+\frac{1}{2}d''_jT_b^2)-\alpha T_b}\\
    &+ \sqrt{\frac{\pi}{d''_j}}e^{-\frac{i}{2}d''_j(s_j-s_0)^2}\left[E(Z_b)-E(Z_a)\right].\nonumber
\end{align}
where $s_0 = s_j - (d'_j+i\alpha)/d''_j$, $Z_a = \sqrt{d''_j/\pi}(s_j-T_a-s_0)$, and $Z_b = \sqrt{d''_j/\pi}(s_j+T_b-s_0)$.

We find that this procedure exhibits good agreement for $d(s_j)\gg 1$ where it is valid to approximate $B(d)$ as a sinusoid with slowly varying amplitude. However, for the situations of interest here, $d(s_j)$ is frequently small so that this approximation breaks down.
%\rublue{
Thus, we choose not to use Eq.~\ref{eq:SM:D_uniform} for the numerical simulations presented in the main text, but note it may be possible to extend these results to small $d(s_j)$ in the future.
%}
%
Also note that for $n=1$ spatial dimensions, $B(d)=C_0\sin(d)$, so that this procedure always works and provides the correct uniform asymptotic approximation for the jerking and stationary point forces for a wave-dressed active particle interacting with a one dimensional wave-field.
%
%\rured{\footnotesize\sf
%\\\textbf{Yes :-) to the blue.} In fact, as I mention somewhere else, in the process of trying to understand the sections I did not, I came up with an idea that does away with the process of introducing discontinuities. Then the contributions from the jerking points involve an integral that needs to be approximated. The simplest thing would be "piece-wise constant", which collapses the scheme on what we are doing in the main text. But one can do better, and what is proposed here goes along "doing better". Except that having an actual formula one wants to approximate may lead to better results. OK, this is too cryptic. If Pedro does not strangle me, I will explain it at some point. :-)}

\vskip6pt

\enlargethispage{20pt}

%%%%%%%%%% Insert bibliography here %%%%%%%%%%%%%%

\vskip2pc

\bibliographystyle{RS} %%%% .BST file

\bibliography{bibliography} %%%%% .Bib file